\documentclass[a4,11pt]{cip-submit-2019}  
\usepackage{color,soul}
\usepackage{afterpage}

\usepackage{wrapfig}
\usepackage{wasysym}
\usepackage{enumitem}
\usepackage{adjustbox}
\usepackage{ragged2e}
\usepackage[svgnames,table]{xcolor}
\usepackage{tikz}
\usepackage{longtable}
\usepackage{changepage}
\usepackage{setspace}
\usepackage{hhline}
\usepackage{multicol}
\usepackage{tabto}
\usepackage{float}
\usepackage{multirow}
\usepackage{makecell}
\usepackage{mathptmx}
\DeclareSymbolFont{epsilon}{OML}{cmm}{m}{it}
\DeclareMathSymbol{\epsilon}{\mathord}{epsilon}{"0F}
\def\DD{D\kern-.7em\raise0.25ex\hbox{\char '55}\kern.33em}
\usepackage{hyperref}
\hypersetup{
	colorlinks=false,
	pdfborder={0 0 0}
}
\usepackage{graphicx,subfigure,wrapfig,epstopdf}
\def\Mr{\uppercase}
\usepackage{graphics}
\usepackage{wasysym}
\usepackage{amsmath,amsxtra,amssymb,latexsym,amscd,color,cite,bm}
\usepackage[mathscr]{eucal}
\usepackage{array}
\usepackage{multirow}
\usepackage{cite} 

\def\doi#1{\href{http://dx.doi.org/#1}{doi:#1}}
\def\titles#1{\title{\large\bf\noindent #1}}
\def\authors#1{\author{\begin{flushleft}{#1}\end{flushleft}}}
\def\authord#1#2{\indent\Mr{#1}$^{#2}$}
\def\addressed#1#2{\\[1mm]\textit{$\!\!\!^{#1}$\indent#2}}

\def\Email{$^{\dagger}$}
\def\email#1{\bigskip\href{mailto:#1}{\!\!\Email\textit{E-mail:}~{#1}}\\[3mm]}
\def\PublicationInformation#1#2#3#4{\\[4mm]\href{mailto:#1}{\!\!\Email\textit{E-mail:}~{#1}}\\[3mm]
	\textit{\indent Received #2}\\[1mm]
	\textit{Accepted for publication~#3}\\[1mm]
	\textit{Published~#4}}
\def\Keywords#1{\\[.2cm] \textnormal{Keywords:~{#1}}.} 
\def\AND{$\text{\Small AND }$}
\def\and{$\text{\tiny AND }$}

\def\Classification#1{\\[.2cm] \textnormal{Classification numbers:~{#1}.}}  

\newcommand{\sout}[1]{\unskip}

\def\ed{
	\bibliographystyle{cip-sty-2019}
	\bibliography{references-database-name}
\end{document}
}
\begin{document}
	\Year{2022}
	\Page{1}\Endpage{15}
	\titles{Anisotropic constant-roll {\it k}-inflation model}
	\authors{
	\authord{Duy H. Nguyen}{1,2}, \authord{Tuyen M. Pham}{1,2}, \authord{Thien D. Le}{3} \\ \AND \authord{Tuan Q. Do}{1,4}\Email
	\newline
	\addressed{1}{Phenikaa Institute for Advanced Study, Phenikaa University, Hanoi 12116, Vietnam}
	\addressed{2}{Graduate University of Science and Technology, Vietnam Academy of Science and Technology, Hanoi 11307, Vietnam}
	\addressed{3}{Department of Natural Science, Phu Cuong High School, Hoa Binh 36100, Vietnam}
	\addressed{4}{Faculty of Basic Sciences, Phenikaa University, Hanoi 12116, Vietnam}
\PublicationInformation{tuan.doquoc@phenikaa-uni.edu.vn}{xxx}{xxx}{xxx}
	}
	\maketitle
	\markboth{Duy H. Nguyen, Tuyen M. Pham, Thien D. Le, and Tuan Q. Do}{Anisotropic constant-roll {\it k}-inflation model}

\begin{abstract}
In this paper, we would like to figure out whether a {\it k}-inflation model admits the Bianchi type I metric as its inflationary solution under a constant-roll condition in the presence of the supergravity motivated coupling between scalar and vector fields, $f^2(\phi)F_{\mu\nu}F^{\mu\nu}$. As a result, some novel anisotropic inflationary solutions are shown to appear along with a power-law one in this scenario. Furthermore, these solutions are numerically confirmed to be attractive, in contrast to the prediction of the Hawking's cosmic no-hair conjecture.
\Classification{98.80.-k, 98.80.Cq, 98.80.Jk
}
\end{abstract} 

\section{\Mr{Introduction}} \label{intro}
Recently, non-trivial anomalies in the cosmic microwave background (CMB) such as the hemispherical asymmetry and the cold spot, which was firstly detected by the Wilkinson Microwave Anisotropy Probe (WMAP) satellite \cite{WMAP} and then confirmed by the Planck one \cite{Planck}, have emerged as one of the leading challenges to the validity of the cosmological principle, which states that the universe should be homogeneous and isotropic on large scales  \cite{cosmological-principle}. It should be noted that the Friedmann-Lemaitre-Robertson-Walker (FLRW) spacetime has been well known as the simplest metric obeying this principle \cite{FLRW}. Therefore, using the homogeneous but anisotropic metrics such as the Bianchi spaces \cite{bianchi} instead of the FLRW one for the early universe might be a reasonable resolution to the CMB anomalies \cite{Schwarz:2015cma}.  It is worth noting that a recent study has pointed out that the Hubble tension might be a signal for a breakdown of the FLRW cosmology \cite{Krishnan:2021dyb}. 

If the early universe was anisotropic, e.g., as described by the Bianchi spaces, non-trivial effects of spatial anisotropies during an inflationary phase \cite{cosmic-inflation} might be imprinted in the CMB \cite{Pitrou:2008gk}. One then could ask a basic and important question that what would a late time state of the universe be? It turns out that this question is not easy to answer. Observationally, it is worth noting that a recent paper has provided a smoking-gun evidence that the current universe might be anisotropic rather than isotropic \cite{Colin:2018ghy}. Theoretically, this question could be answered according to the so-called cosmic no-hair conjecture proposed by Hawking and his colleagues several decades ago \cite{cosmic-no-hair}. In particular, if this conjecture was valid, the late time universe would simply be homogeneous and isotropic on large scales, i.e., would obey the cosmological principle. Unfortunately, a general proof for this conjecture has remained unknown for several decades. It appears that many people have made great efforts to archive partial proofs for this conjecture since the Wald's seminal one \cite{wald,proof-1,proof-2,proof-3}. For a very recent extension of the Wald's proof, see Ref. \cite{Azhar:2022yip}. Some other important discussions on this conjecture can be found in papers \cite{local-1,local-2}. In particular, these papers have pointed out that  the cosmic no-hair conjecture would only be valid locally, i.e., inside of the future event horizon.  Along with the partial proofs, some counterexamples to this conjecture have been claimed to appear in several cosmological models. However, many of them have been shown to be invalid due to instabilities. Recently, a very truly counterexample to this conjecture has been found in a supergravity-motivated model by Kanno, Soda, and Watanabe (KSW) \cite{MW0,MW}. As a result, this model admits a stable and attractive Bianchi type I solution during an inflationary phase due to the existence of unusual coupling between scalar and vector fields, i.e., $f^2(\phi)F_{\mu\nu}F^{\mu\nu}$. It turns out that the existence of non-trivial gauge kinetic function $f(\phi)$ prevents the vector field from a rapid dilution during the inflationary phase and therefore leads to stable spatial anisotropies (a.k.a. spatial hairs) of spacetime. Consequently, many extended counterexamples have been figured out in a number of follow-up papers, e.g., see an incomplete list in Refs. \cite{extensions,Do:2011zz,Ohashi:2013pca,Do:2020hjf,Ito:2017bnn,Nguyen:2021emx}. Many other cosmological aspects of the KSW model can be found in interesting review papers \cite{Maleknejad:2012fw}.

 Currently, we are interested in investigating two non-trivial extensions of the KSW model. The first one is non-canonical models, in which a canonical scalar field considered in the KSW model is replaced by a non-canonical one such as  the string-inspired Dirac-Born-Infeld field \cite{Do:2011zz} or the {\it k}-inflation field \cite{Ohashi:2013pca,Do:2020hjf}. As a result, these non-canonical models have been shown to admit stable and attractive anisotropic power-law inflationary solutions, which do violate the cosmic no-hair conjecture. The latter one is novel types of anisotropic inflation \cite{Ito:2017bnn,Nguyen:2021emx}, which have been investigated under the so-called constant-roll condition \cite{Motohashi:2014ppa}. See also Refs. \cite{constant-roll-extension,Odintsov:2019ahz} for various interesting discussions on this constant-roll inflation. As mentioned in Ref. \cite{Motohashi:2014ppa}, the constant-roll inflation has been proposed as an interpolation between the well-known slow-roll inflation and a novel ultra slow-roll inflation proposed in Ref. \cite{Martin:2012pe}. It appears that the constant-roll condition, in which one of the slow-roll parameters, $\eta \equiv -\ddot\phi/(H\dot\phi)$, is required to be constant during the inflationary phase, leads to not only the well-known solutions, especially the power-law one previously found  in Refs. \cite{Abbott:1984fp,Barrow:1994nt,Boubekeur:2005zm}, but also novel ones \cite{Motohashi:2014ppa}, whose cosmological consequences might be slightly different from usual ones. 

In this paper, we would like to combine these two extensions in a single scenario, in which the canonical scalar field is replaced by the non-canonical one firstly proposed in the so-called {\it k}-inflation model \cite{ArmendarizPicon:1999rj} and the constant-roll condition,  $\eta \equiv -\ddot\phi/(H\dot\phi)={\text {constant}}$, is applied. As a result, we will figure out analytical inflationary solutions under the constant-roll condition for the {\it k}-inflation. Furthermore, we will show that these solutions are indeed attractive and therefore violate the cosmic no-hair conjecture.  It should be noted that Ref. \cite{Odintsov:2019ahz} also considers the {\it k}-inflation under the constant-roll condition for the FLRW metric. However, it does not include the effect of vector field. Consequently, its solutions are different from that derived in this paper.

As a result, this paper will be organized as follows: (i) An introduction has been written in Sec. I. A basic setup of the proposed model will be shown in Sec. II. Then, analytical solutions under the constant-roll condition will be solved and presented in Sec. III. The attractive behavior during the inflationary phase of the obtained solutions will be numerically confirmed in Sec. IV. Finally,  concluding remarks will be written in Sec. V.
\section{\Mr{Basic setup}} \label{chap2}
Let us start with a general action for the {\it k}-inflation extension of the KSW model  given by \cite{Ohashi:2013pca,Do:2020hjf}
\begin{align}
S=\int d^4x\sqrt{-g}\left[\frac{R}{2}+P(X,\phi)-\frac{1}{4}f^2(\phi)F_{\mu\nu}F^{\mu\nu}\right],
\end{align}
where the reduced Planck mass, $M_p$, is set to be one for convenience. Additionally,  $P(\phi,X)$ is an arbitrary function of scalar field $\phi$ and its kinetic $X\equiv-\partial^\mu\phi\partial_\mu\phi/2$ \cite{ArmendarizPicon:1999rj}, while $F_{\mu\nu}\equiv \partial_\mu A_\nu -\partial_\nu A_\mu$ is the field strength of vector field $A_\mu$.  It is noted again that the last term in the above action, which is a supergravity-motivated coupling, is the key ingredient of the KSW model  \cite{MW0,MW}. More specifically, this coupling can be seen in the bosonic part of the action of four-dimensional  $\mathcal {N}=1$ supergravity as mentioned in Ref. \cite{MW}. The reason why the bosonic part of supergravity is more relevant to the inflationary phase than the fermionic part is that the origin of inflaton, which is a scalar field responsible for causing the inflationary phase, could be realized in this part. One might, however, ask that if fermions are not important to the inflationary phase, why should we consider the supergravity, which involves not only bosons but also fermions, for this phase? In principle, one could propose inflationary models due to specific physical reasons, typically the invariance of action under field transformations.  Of course, their predictions should be consistent with the observational data. If successful, realizing the proposed inflationary models in the context of the underlying theories such as the supergravity, which have been assumed to take place in high energy levels of the early universe, would be the next necessary step. Unfortunately, this task is not straightforward at all. For example, many people have tried to realize the Starobinsky model \cite{cosmic-inflation}, which has been one of the most viable inflationary models, in the $\mathcal {N}=1$ supergravity \cite{Ellis:2013xoa,Farakos:2013cqa}. Therefore, figuring out inflationary models within the framework of supergravity seems to be a natural approach \cite{Yamaguchi:2011kg}. Moreover, the supergravity could provide us extra unusual couplings between boson fields, e.g., the coupling between scalar and vector fields shown above, which might affect on the dynamics of the inflationary phase and therefore might lead to unexpected consequences \cite{MW}. More interestingly, they might be reasonable resolutions to the remaining problems in cosmology such as the CMB anomalies \cite{Schwarz:2015cma}.  As a result, the existence of the gauge kinetic function $f(\phi)$ breaks down the conformal invariance of vector field and therefore prevents it from a rapid dilution during the inflationary phase. Consequently, it would lead to a stable anisotropic inflation violating the prediction of the cosmic no-hair conjecture. 

In this paper, following both Refs. \cite{Do:2020hjf} and \cite{Odintsov:2019ahz} we will choose the form of $P(X,\phi)$ as follows
\begin{align}
P(X,\phi)=X+L(\phi)X^2-V(\phi),
\end{align}
where $L(\phi)$ is an undetermined function of  the scalar field $\phi$ and $V(\phi)$ is its potential. It should be noted that this form of $P(X,\phi)$ is different from that given in Ref. \cite{Do:2020hjf} as well as that given in Ref. \cite{Odintsov:2019ahz}. In particular, $P(X,\phi)$ in 
Ref. \cite{Do:2020hjf} does not include the potential $V(\phi)$, while $L(\phi)$ in 
Ref. \cite{Odintsov:2019ahz} has been set to be constant. Additionally, $\phi$ will only have a positive kinetic energy.  As a result, $V(\phi)$ should not be removed in the presence of the constant-roll condition, while the existence of $L(\phi)$ is necessary for having power-law solutions of scale factors. It appears that this proposed model will reduce to the KSW model of canonical scalar field \cite{MW0,MW,Ito:2017bnn} once $L(\phi)$ vanishes.

We will consider, according to the previous works in Refs. \cite{Ito:2017bnn,Nguyen:2021emx}, the homogeneous but anisotropic Bianchi type I metric as 
\begin{align}
ds^2=-dt^2+a^2(t)\left[b^{-4}(t)dx^2+b^2(t)(dy^2+dz^2)\right],
\end{align}
where $a(t)$ plays as an isotropic scale factor, while $b(t)$ acts as a deviation from isotropy.  In addition, the scalar field will be assumed to be homogeneous, i.e., $\phi=\phi(t)$, while the vector field will be taken as $A_\mu(t)=(0,A_x(t),0,0)$ in order to be compatible with the Bianchi type I metric. 

As a result, a non-trivial solution of the vector field can be solved to be
\begin{align}
\dot{A}_x=p_A f^{-2}a^{-1}b^{-4},
\end{align}
with $p_A$ is an integration constant \cite{Ito:2017bnn,Nguyen:2021emx}. Note that $\dot A_x \equiv dA_x/dt$. Consequently, the corresponding Einstein field equations turn out to be
\begin{align}
\label{1st Einstein equation}
H_a^2-H_b^2&=\frac{1}{6}\dot{\phi}^2+\frac{L}{4}\dot{\phi}^4+\frac{V}{3}+\frac{f^{-2}}{6}a^{-4}b^{-4}p_A^2, \\
 \label{2nd Einstein equation}
\dot{H}_a+3H_b^2&=-\frac{1}{2}\dot{\phi}^2-\frac{1}{2}L\dot{\phi}^4-\frac{f^{-2}}{3}a^{-4}b^{-4}p_A^2,\\
\label{3rd Einstein equation}
\dot{H}_b+3H_aH_b&=\frac{f^{-2}}{3}a^{-4}b^{-4}p_A^2,
\end{align}
along with  the scalar field equation given by
\begin{align}
\label{scalar field equation}
\left(1+3L\dot{\phi}^2 \right)\ddot{\phi}=-3 \left(1+L\dot{\phi}^2 \right)H_a\dot{\phi}-V'-\frac{3}{4}L'\dot{\phi}^4+f^{-3}f'a^{-4}b^{-4}p_A^2.
\end{align}
Here $H_a \equiv \dot a/a$ and $H_b \equiv \dot b/b$ acting as the Hubble parameters. Additionally, $f'\equiv \partial f/\partial \phi$ and $V' \equiv \partial V/\partial \phi$.
It is straightforward to see that if we set $L=0$ then all these equations will reduce to that derived in Ref. \cite{Ito:2017bnn} for the canonical scalar field.
\section{\Mr{Solutions under a constant-roll condition}} \label{chap3}
Given the field equations derived in the previous section, we would like to seek anisotropic solutions under the constant-roll condition \cite{Motohashi:2014ppa,Ito:2017bnn,Nguyen:2021emx},
\begin{align}
\eta(t)\equiv-\frac{\ddot{\phi}}{H_a\dot{\phi}}=\hat{\beta},\label{constant-roll condition}
\end{align}
along with the constant anisotropy condition \cite{Ito:2017bnn,Nguyen:2021emx},
\begin{align}
\frac{H_b}{H_a}=n,\label{constant anisotropy condition}
\end{align}
where $\hat{\beta}$ and $n$ are all constant. As a result, Eqs. (\ref{2nd Einstein equation}), (\ref{3rd Einstein equation}), and (\ref{constant anisotropy condition}) lead to 
\begin{align}
(1+n)\dot{H}_a+3n(1+n)H_a^2+X+2LX^2=0. \label{eq1}
\end{align}
Due to the non-canonical property of scalar field, we introduce its speed of sound parameter as follows \cite{ArmendarizPicon:1999rj}
\begin{align}
c_s^2=\frac{\partial_X p}{\partial_X \rho}=\frac{1+2LX}{1+6LX},
\end{align}
where $p$ and $\rho$ are pressure and density parameters given by
\begin{align}
p&=P(\phi,X)=X+LX^2-V,\\
\rho&=2X \partial_X P(\phi,X)-P(\phi,X)=X+3LX^2+V,
\end{align}
respectively.
Here $\partial_X p \equiv \partial p /\partial X$. It turns out that $c_s=1$ for the canonical scalar field, while $c_s <1$ for the non-canonical one with a positive $L$. It should be noted that our previous paper is also on a non-canonical scalar field model, which is nothing but the Dirac-Born-Infeld (DBI) model with $c_s = \gamma^{-1}$ and $\gamma >1$ being the Lorentz factor \cite{Nguyen:2021emx}. More importantly, the speed of sound, or equivalently the Lorentz factor, has been assumed to be constant in order to have  an anisotropic constant-roll inflation in this DBI model. In this paper, therefore, we will also assume that 
\begin{align}
c_s=\hat{c}_s, \label{constant speed of sound condition}
\end{align}
where $0<\hat{c}_s \leq 1$ is a constant. The constancy of the speed of sound $c_s$ seems to be an additional important condition for having a constant-roll inflation in non-canonical scalar field models. 

For convenience, we rewrite Eq. (\ref{eq1}) as 
\begin{align}
(1+n)\dot{H}_a+3n(1+n)H_a^2+\hat{c}_s^2 X+6\hat{c}_s^2 LX^2=0, \label{eq2}
\end{align}
which will become as
\begin{align}
\hat{c}_s^2 \dot{\phi}^2+(3\hat{c}_s^2-1)(1+n)H_a'\dot{\phi}+(9\hat{c}_s^2-3)n(1+n)H_a^2=0,
\end{align}
if $H_a$ is regarded as a function of $\phi$. As a result, solving this equation gives non-trivial solutions,
\begin{align}
\dot{\phi}=\frac{(1-3\hat{c}_s^2)H_a'(1+n)\pm\sqrt{(1+n)^2(3\hat{c}_s^2-1)^2 H_a'^2-12\hat{c}_s^2(3\hat{c}_s^2-1)n(1+n) H_a^2}}{2 \hat{c}_s^2}. \label{solution for d phi/dt}
\end{align}
Furthermore, differentiating these solutions with respect to $t$ implies that
\begin{align} \label{key-equation}
-\hat{\beta} H_a=\frac{(1-3\hat{c}_s^2)(1+n)H_a''}{2\hat{c}_s^2 }\pm\frac{\sqrt{1+n}\left[(1+n)(3\hat{c}_s^2-1)^2 H_a' H_a''-12\hat{c}_s^2(3\hat{c}_s^2-1)n H_a H_a'\right]}{2 \hat{c}_s^2 \sqrt{(1+n)(3\hat{c}_s^2-1)^2 H_a'^2-12\hat{c}_s^2(3\hat{c}_s^2-1)n H_a^2}},
\end{align}
here $H''_a \equiv d^2 H_a/d\phi^2$. Now, we consider the isotropic case with $n=0$. It appears that the vanishing of $n$ leads Eq. \eqref{key-equation} to 
\begin{align}
\frac{(3\hat{c}_s^2-1)H_a''}{\hat{c}_s^2}=\hat{\beta} H_a,
\end{align}
which admits a general solution,
\begin{align}
H_a(\phi)=C_1 \exp\left(\sqrt{\frac{\hat{\beta}}{3\hat{c}_s^2-1}}\hat{c}_s\phi\right)+C_2  \exp\left(-\sqrt{\frac{\hat{\beta}}{3\hat{c}_s^2-1}}\hat{c}_s\phi\right),
\end{align}
where $C_1$ and $C_2$ are nothing but arbitrary integration constants.
Hinted by this solution and that in Refs. \cite{Ito:2017bnn,Nguyen:2021emx}, we assume that a general solution of Eq. \eqref{key-equation} with the non-vanishing $n$ will take the following ansatz,
\begin{align}
H_a(\phi)&=C_1 \exp\left(\lambda(n)\sqrt{\frac{\hat{\beta}}{3\hat{c}_s^2-1}}\hat{c}_s\phi\right)+C_2  \exp\left(-\lambda(n)\sqrt{\frac{\hat{\beta}}{3\hat{c}_s^2-1}}\hat{c}_s\phi\right),
\end{align}
here $\lambda(n)$ is a non-trivial function of $n$. As a result, we are able to figure out the desired general solution of  $H_a$ as
\begin{align}
H_a(\phi)=C_1 \exp\left(\sqrt{\frac{12\hat{\beta}}{(3\hat{c}_s^2-1)(\hat{\beta}+6)}}\hat{c}_s\phi\right)+C_2 \exp\left(-\sqrt{\frac{12\hat{\beta}}{(3\hat{c}_s^2-1)(\hat{\beta}+6)}}\hat{c}_s\phi\right), \label{general solution of H_a}
\end{align}
provided a relation 
\begin{align}
n=\frac{\hat{\beta}}{6},
\end{align}
which was firstly used in Ref. \cite{Ito:2017bnn} and then re-used in Ref. \cite{Nguyen:2021emx} in order to have the corresponding anisotropic constant-roll solutions. This is indeed a necessary condition for having the anisotropic constant-roll inflation. Consequently, the corresponding general solution for $H_b$ is given by
\begin{align}
H_a(\phi)=C_1\frac{\hat{\beta}}{6} \exp\left(\sqrt{\frac{12\hat{\beta}}{(3\hat{c}_s^2-1)(\hat{\beta}+6)}}\hat{c}_s\phi\right)+C_2 \frac{\hat{\beta}}{6}\exp\left(-\sqrt{\frac{12\hat{\beta}}{(3\hat{c}_s^2-1)(\hat{\beta}+6)}}\hat{c}_s\phi\right).\label{general solution of H_b}
\end{align}
In the next subsections, we will discuss, following the analysis in Refs. \cite{Ito:2017bnn,Nguyen:2021emx,Motohashi:2014ppa}, whether the anisotropic constant-roll inflation exists in three different specific cases of $C_1$ and $C_2$.
\subsection{First solution}\label{first solution}
In this subsection, we would like to consider the case with $C_1=M$ and $C_2=0$, where $M$ is an integration constant determining the amplitude
of the power spectrum of the curvature perturbation \cite{Motohashi:2014ppa}. As a result, the corresponding $H_a$ and $H_b$ become as
\begin{align}
H_a(\phi)&=M \exp\left(\sqrt{\frac{12\hat{\beta}}{(3\hat{c}_s^2-1)(\hat{\beta}+6)}}\hat{c}_s\phi\right), \label{Ha exp}\\
H_b(\phi)&=\frac{M\hat{\beta}}{6} \exp\left(\sqrt{\frac{12\hat{\beta}}{(3\hat{c}_s^2-1)(\hat{\beta}+6)}}\hat{c}_s\phi\right). \label{Hb exp}
\end{align}
It appears that  $H_a$ and $H_b$ will  be real definite if $c_s^2>1/3$, provided that $\hat{\beta}>0$, i.e,  $n>0$. Hence, the corresponding $\dot{\phi}$ can be solved to be
\begin{align}
\dot{\phi}=-\frac{M\hat{\beta}}{\sqrt{\frac{12 \hat{\beta}}{(3\hat{c}_s^2-1)(6+\hat{\beta})}}\hat{c}_s} \exp\left(\sqrt{\frac{12 \hat{\beta}}{(3\hat{c}_s^2-1)(6+\hat{\beta})}}\hat{c}_s\phi\right), \label{dphi/dt exp}
\end{align}
with the help of \eqref{solution for d phi/dt}. Thanks to these results, we are able to derive the following equation for $a$ as
\begin{align}
\frac{a'}{a}=-\frac{\hat{c}_s}{\hat{\beta}}\sqrt{\frac{12 \hat{\beta}}{(3\hat{c}_s^2-1)(6+\hat{\beta})}}, \label{eq of a exp}
\end{align} 
along with that for $b$ given by
\begin{align}
\frac{b'}{b}=-\frac{\hat{c}_s}{6}\sqrt{\frac{12 \hat{\beta}}{(3\hat{c}_s^2-1)(6+\hat{\beta})}}, \label{eq of b exp}
\end{align}
here $a' \equiv da/d\phi$ and $b' \equiv db/d\phi$. 
Consequently,  the scale factors $a$ and $b$ can be expressed as a function, up to a constant, of the scalar field as
\begin{align}
a&\propto \exp\left(-\frac{\hat{c}_s}{\hat{\beta}}\sqrt{\frac{12 \hat{\beta}}{(3\hat{c}_s^2-1)(6+\hat{\beta})}}\phi\right), \label{a exp}\\
b&\propto \exp\left(-\frac{\hat{c}_s}{6}\sqrt{\frac{12 \hat{\beta}}{(3\hat{c}_s^2-1)(6+\hat{\beta})}}\phi\right), \label{b exp}
\end{align}
respectively. Using Eqs. \eqref{1st Einstein equation}, \eqref{2nd Einstein equation}, and \eqref{3rd Einstein equation}, we are able to determine the corresponding form of  $f(\phi)$, $L(\phi)$, and $V(\phi)$ as
\begin{align}
f(\phi)&\propto \exp\left( \sqrt{\frac{12 \hat{\beta}}{(3\hat{c}_s^2-1)(6+\hat{\beta})}}\frac{\hat{c}_s(6-2\hat{\beta})}{3\hat{\beta}}\phi\right), \label{f exp}\\
L(\phi)&=\frac{-12 \hat{c}_s^2(\hat{c}_s^2-1)}{(1-3\hat{c}_s^2)^2 M^2 \hat{\beta}(6+\hat{\beta})}\exp\left(-2\sqrt{\frac{12\hat{\beta}}{(3\hat{c}_s^2-1)(6+\hat{\beta})}}\hat{c}_s\phi\right), \label{L exp}\\
V(\phi)&=\frac{M^2\left[\hat{c}_s^2(\hat{\beta}-6)(5\hat{\beta}-24)-\hat{\beta}(\hat{\beta}+6)\right]}{48\hat{c}_s^2}\exp\left(2\sqrt{\frac{12\hat{\beta}}{(3\hat{c}_s^2-1)(6+\hat{\beta})}}\hat{c}_s\phi\right),\label{V exp}
\end{align}
respectively. It should be noted that if we solve Eq. \eqref{dphi/dt exp}, the explicit form of $\phi$ will be shown to be
\begin{equation}
\phi =-\frac{1}{\hat{c}_s}\sqrt{\frac{(3\hat{c}_s^2-1)(6+\hat{\beta})}{12\hat{\beta}}}\log\left(M\hat{\beta}t \right).
\end{equation}
More interestingly, inserting this solution into Eqs. \eqref{a exp} and \eqref{b exp} leads to the corresponding power-law solution, up to a constant,
\begin{align}
\label{a-power}
a(t) &\propto t^{1/\hat{\beta}},\\
\label{b-power}
b(t)&\propto t^{1/6}.
\end{align} 
It turns out that $0<\hat\beta\ll 1$ is a sufficient condition for having an anisotropic power-law inflation with small anisotropies.  
\subsection{Second solution}\label{second solution}
In this subsection, we would like to deal with the case, in which  $C_1=C_2=M/2$.  As a result, the corresponding Hubble parameters, $H_a$ and $H_b$, become as
\begin{align}
H_a(\phi)&=M \cosh\left(\sqrt{\frac{12\hat{\beta}}{(3\hat{c}_s^2-1)(\hat{\beta}+6)}}\hat{c}_s\phi\right), \\
H_b(\phi)&=\frac{M\hat{\beta}}{6} \cosh\left(\sqrt{\frac{12\hat{\beta}}{(3\hat{c}_s^2-1)(\hat{\beta}+6)}}\hat{c}_s\phi\right),
\end{align}
respectively. It is important to note that this solution has been shown to be not attractive for the canonical scalar field with $\hat c_s=1$ \cite{Ito:2017bnn}. This is also the case for this {\it k}-inflation model with $1/3<\hat c_s<1$, so we will ignore this solution from now on. 
\subsection{Third solution}\label{third solution}
In this subsection, we would like to consider the case, in which $C_1=M/2$ and $C_2=-M/2$.  As a result, the corresponding Hubble parameters, $H_a$ and $H_b$, read 
\begin{align}
H_a(\phi)&=M \sinh\left(\sqrt{\frac{12\hat{\beta}}{(3\hat{c}_s^2-1)(\hat{\beta}+6)}}\hat{c}_s\phi\right), \label{Ha sinh}\\
H_b(\phi)&=\frac{M\hat{\beta}}{6} \sinh\left(\sqrt{\frac{12\hat{\beta}}{(3\hat{c}_s^2-1)(\hat{\beta}+6)}}\hat{c}_s\phi\right),\label{Hb sinh}
\end{align}
respectively. 
Similar to the first solution, we will only consider  $\hat{\beta}>0$, i.e., $n>0$ as well as $\hat c_s^2>1/3$, in order to have real definite $H_a$ and $H_b$. Furthermore, the corresponding solution for $\dot{\phi}$ is solved to be
\begin{align}
\dot{\phi}=\frac{M}{\hat{c}_s}\sqrt{\frac{(3\hat{c}_s^2-1)\hat{\beta}(6+\hat{\beta})}{12}}\left[\pm 1- \cosh\left(\sqrt{\frac{12\hat{\beta}}{(6+\hat{\beta})(3\hat{c}_s^2-1)}}\hat{c}_s\phi\right)\right], \label{dphi/dt sinh}
\end{align}
the help of \eqref{solution for d phi/dt}. Plugging this solution into Eqs. \eqref{Ha sinh} and \eqref{Hb sinh} leads to the corresponding solution for $a$ and $b$, up to a constant, as
\begin{align}
a&\propto\left[\mp 1+\cosh\left(\sqrt{\frac{12\hat{\beta}}{(3\hat{c}_s^2-1)(6+\hat{\beta})}}\hat{c}_s\phi\right)\right]^{-\frac{1}{\hat{\beta}}}, \label{a sinh}\\
b&\propto\left[\mp 1+\cosh\left(\sqrt{\frac{12\hat{\beta}}{(3\hat{c}_s^2-1)(6+\hat{\beta})}}\hat{c}_s\phi\right)\right]^{-\frac{1}{6}},\label{b sinh}
\end{align}
respectively. Consequently, the corresponding forms of $f(\phi)$, $L(\phi)$, and $V(\phi)$ turn out to be
\begin{align}
f(\phi)\propto&~ \frac{\left[\mp 1+\cosh\left(\sqrt{\frac{12\hat{\beta}}{(3\hat{c}_s^2-1)(6+\hat{\beta})}}\hat{c}_s \phi\right)\right]^\frac{\hat{\beta}+6}{3\hat{\beta}}}{\sqrt{-(\hat{\beta}+3)-(\hat{\beta}-3)\cosh\left(2\sqrt{\frac{12\hat{\beta}}{(3\hat{c}_s^2-1)(6+\hat{\beta})}}\hat{c}_s \phi\right)\pm 2\hat{\beta}\cosh\left(\sqrt{\frac{12\hat{\beta}}{(3\hat{c}_s^2-1)(6+\hat{\beta})}}\hat{c}_s \phi\right)}}, \label{f sinh}\\
L(\phi)=&~\frac{-12 \hat{c}_s^2(\hat{c}_s^2-1)}{M^2(1-3\hat{c}_s^2)^2 \hat{\beta}(6+\hat{\beta})\left[\cosh\left(\sqrt{\frac{12\hat{\beta}}{(3\hat{c}_s^2-1)(6+\hat{\beta})}}\hat{c}_s\phi\right)\mp 1\right]^2}, \label{L sinh}
\end{align}
\begin{align}
V(\phi)=&~\frac{M^2}{24 \hat{c}_s^2}\left\lbrace\frac{1}{4}\left[\hat{c}_s^2(\hat{\beta}-6)(7\hat{\beta}+24)-3\hat{\beta}(\hat{\beta}+6)\right]\right.\nonumber\\
&+\frac{1}{4}\left[\hat{c}_s^2(\hat{\beta}-6)(5\hat{\beta}-24)-\hat{\beta}(\hat{\beta}+6)\right]\cosh\left(2\sqrt{\frac{12\hat{\beta}}{(3\hat{c}_s^2-1)(6+\hat{\beta})}}\hat{c}_s\phi\right)\nonumber\\
&\left.\mp\hat{\beta}\left[3\hat{c}_s^2(\hat{\beta}-6)-\hat{\beta}-6\right]\cosh\left(\sqrt{\frac{12\hat{\beta}}{(3\hat{c}_s^2-1)(6+\hat{\beta})}}\hat{c}_s\phi\right)\right\rbrace, \label{V sinh}
\end{align}
respectively.  As a result, the solution for Eq. \eqref{dphi/dt sinh} with the upper sign ``+'' is given by
\begin{align} \label{phi+}
\phi(t)=\phi_+(t)=\sqrt{\frac{(6+\hat{\beta})(3\hat{c}_s^2-1)}{3\hat{\beta}\hat{c}_s^2}}\text{arccoth}\left(M\hat{\beta}t \right),
\end{align}
which leads to the corresponding scale factors defined, up to a constant, as
\begin{align} \label{a-upper-sign}
a(t)&\propto(M^2\hat{\beta}^2t^2-1)^{1/\hat{\beta}}, \\
 \label{b-upper-sign}
b(t)&\propto(M^2\hat{\beta}^2t^2-1)^{1/6}.
\end{align}
On the other hand, Eq. \eqref{dphi/dt sinh} with the lower sign ``-'' admits the corresponding solution,
\begin{align} \label{phi-}
\phi(t)=\phi_-(t)=-\sqrt{\frac{(6+\hat{\beta})(3\hat{c}_s^2-1)}{3\hat{\beta}\hat{c}_s^2}}\text{arctanh} \left(M\hat{\beta}t \right),
\end{align}
which implies the corresponding scale factors defined, up to a constant, as
\begin{align}
 \label{a-lower-sign}
a(t)&\propto(1-M^2\hat{\beta}^2t^2)^{1/\hat{\beta}},\\
 \label{b-lower-sign}
b(t)&\propto(1-M^2\hat{\beta}^2t^2)^{1/6}.
\end{align}
Interestingly, the obtained solutions in this model turn out to be similar, up to a constant, to that found in Ref. \cite{Nguyen:2021emx}. Hence, the discussions in Ref. \cite{Nguyen:2021emx} on whether the corresponding inflation exists are still valid for the solutions derived in this paper. In particular, for the solution shown in Eqs. \eqref{a-upper-sign} and \eqref{b-upper-sign}, the following constraints, $M\hat\beta t >1$, $M>0$, and $0<\hat\beta\ll 1$ need to be fulfilled  for the existence of inflationary phase. On the other hand, the following constraints, $-1<M\hat\beta t <0$, $M<0$, and $0<\hat\beta\ll 1$ are necessary for ensuring that the solution shown in Eqs. \eqref{a-lower-sign} and \eqref{b-lower-sign} will act as an inflationary one. It is important to note that these two inflationary solutions all need to finish before $\phi$ becomes zero due to either the strong coupling problem (associated with $\phi_+$) or the negativity of $V(\phi)$ (associated with $\phi_-$). For detailed explanations, see Ref. \cite{Nguyen:2021emx}. An additional important point should be mentioned is that the $L(\phi)$ obtained in the above cases all turn out to be positive definite, leading to the subluminality of speed of sound, i.e., $c_s(t) <1$.
\section{\Mr{Attractor property}} \label{chap4}
So far, we have derived three different analytical solutions: solution I described by a set of Eqs. \eqref{Ha exp}-\eqref{b-power}; solution II described by a set of Eqs. \eqref{f sinh}, \eqref{L sinh}, \eqref{V sinh}, \eqref{a-upper-sign}, and \eqref{b-upper-sign} associated with the $\phi_+$ solution shown in Eq. \eqref{phi+}; and solution III described by a set of Eqs. \eqref{f sinh}, \eqref{L sinh}, \eqref{V sinh}, \eqref{a-lower-sign}, and \eqref{b-lower-sign} associated with the $\phi_-$ solution shown in Eq. \eqref{phi-}.

\begin{figure}[hbtp]
\includegraphics[scale=0.24]{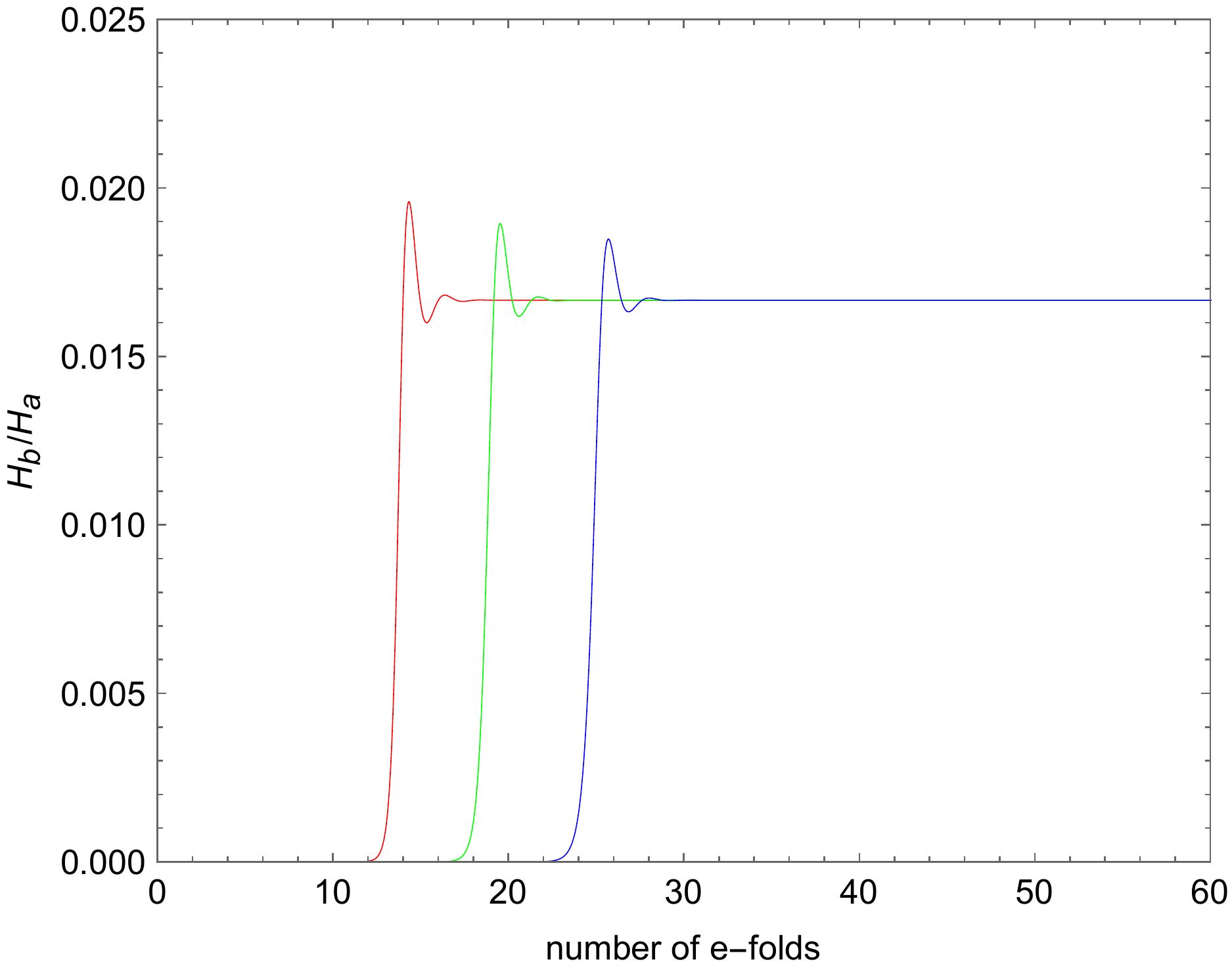}\quad
\includegraphics[scale=0.24]{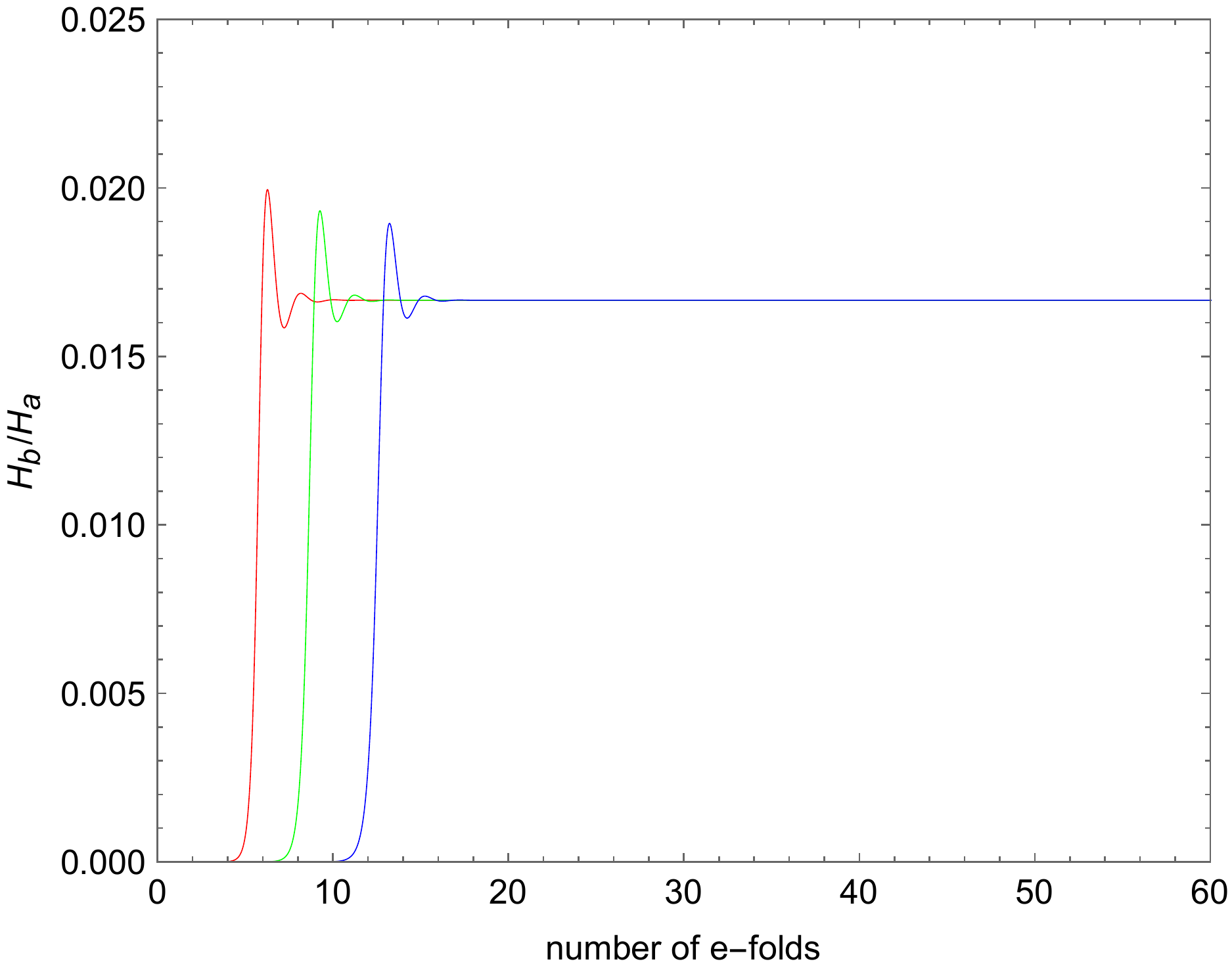}\quad
\includegraphics[scale=0.24]{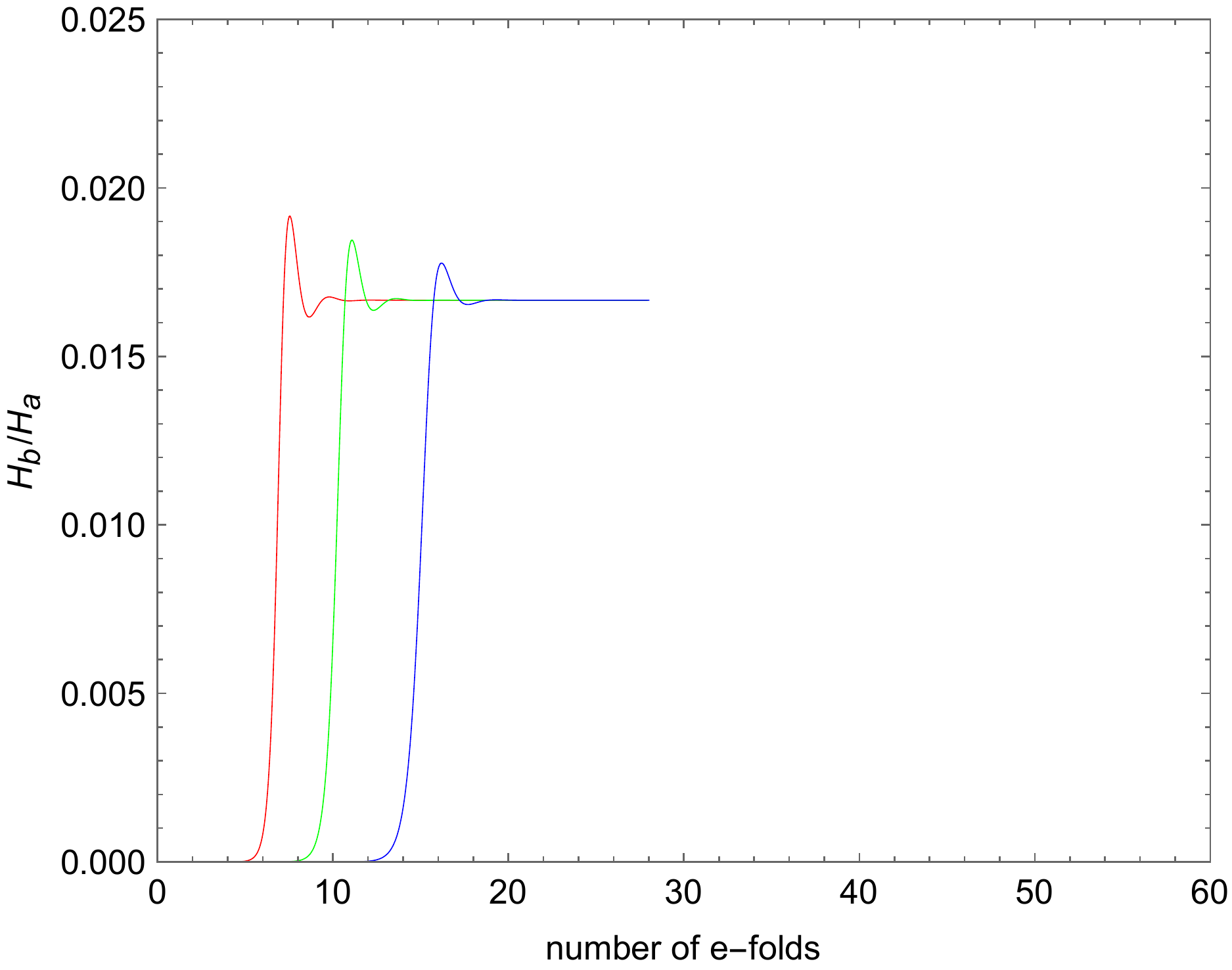}\\
\centering
\caption{Time evolution of the ratio $H_b(t)/H_a(t)$ for three different values of $\hat{c}_s$. It is clear that all trajectories converge to the same constant $n=\hat\beta/6 \simeq 0.0167$, confirming the attractive property $H_b(t)/H_a(t) \to n=\hat\beta/6 \simeq 0.0167$ when the e-fold number increases. The left, middle, and right plots correspond to the solutions {I}, {II}, and {III}, respectively. The red, green, and blue curves correspond to $\hat{c}_s=1,~0.95$, and $0.9$, respectively.}
\label{evolution of n}
\end{figure}

\begin{figure}[hbtp]
\includegraphics[scale=0.24]{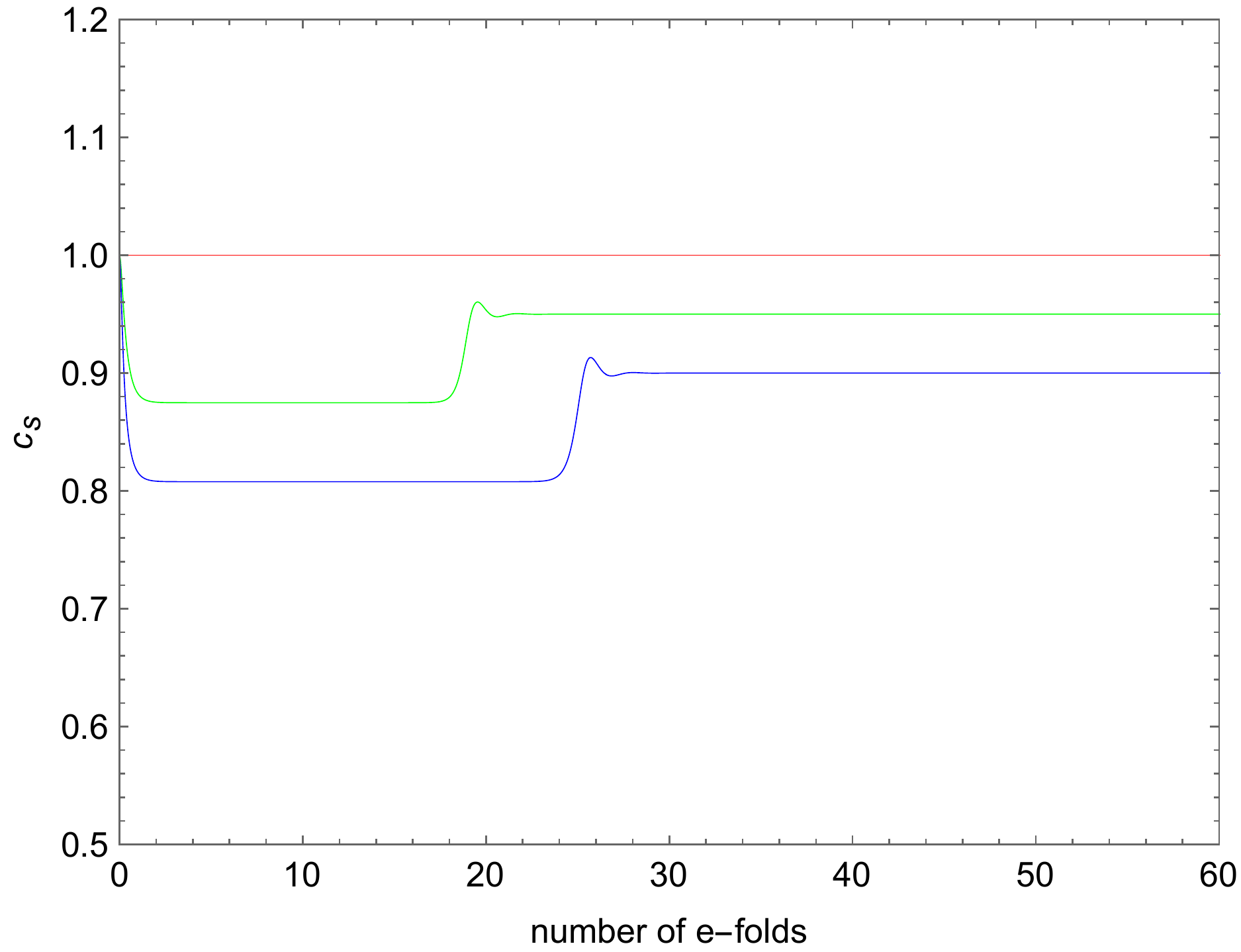}\quad
\includegraphics[scale=0.24]{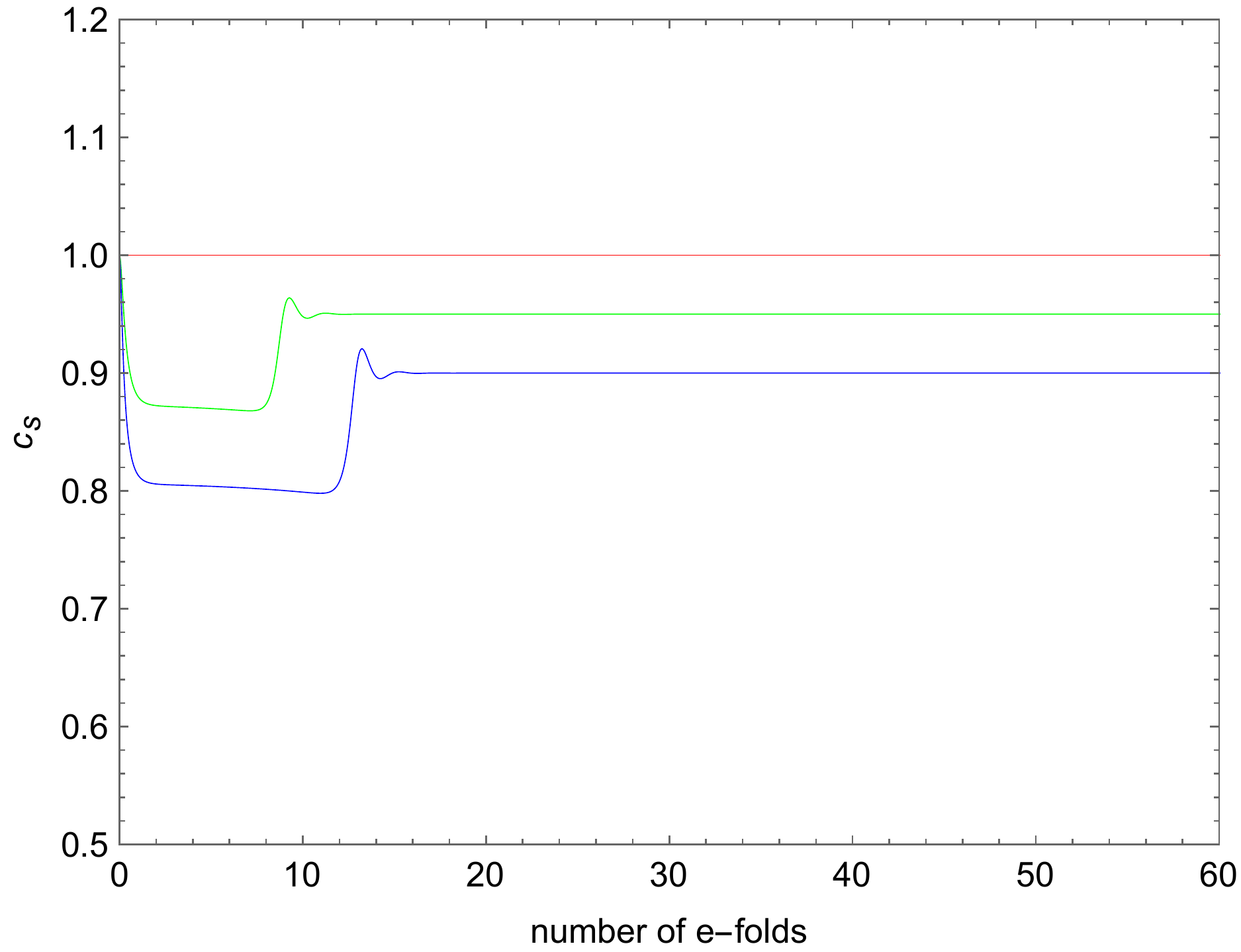}\quad
\includegraphics[scale=0.24]{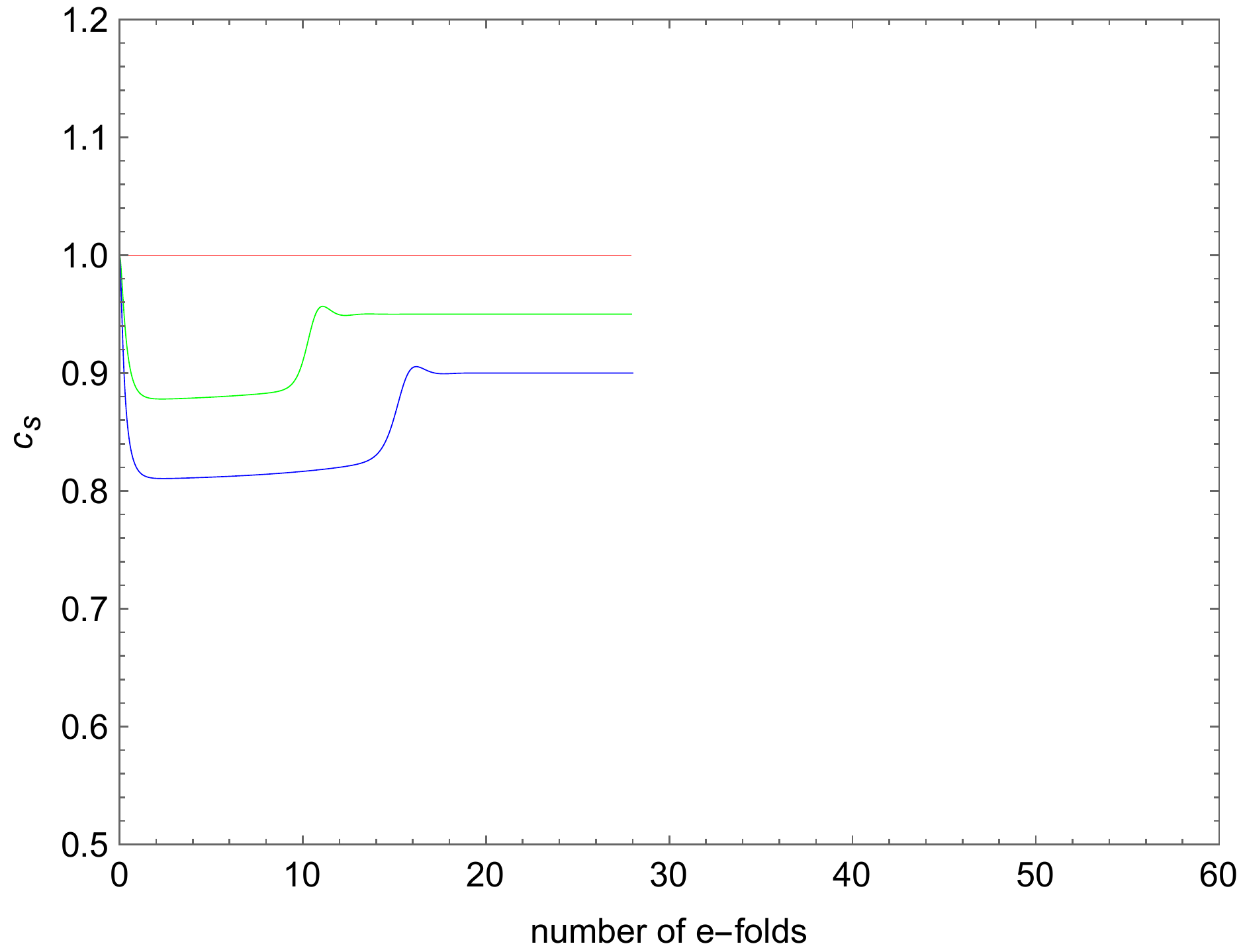}\\
\centering
\caption{Time evolution of $c_s(t)$ for three different values of $\hat{c}_s$. It is clear that  $c_s(t)\to \hat c_s$ when the e-fold number increases. The left, middle, and right plots correspond to the solutions {I}, {II}, and {III}, respectively. The red, green, and blue curves correspond to $\hat{c}_s=1,~0.95$, and $0.9$, respectively.}
\label{evolution of cs}
\end{figure}

\begin{figure}[hbtp]
\includegraphics[scale=0.24]{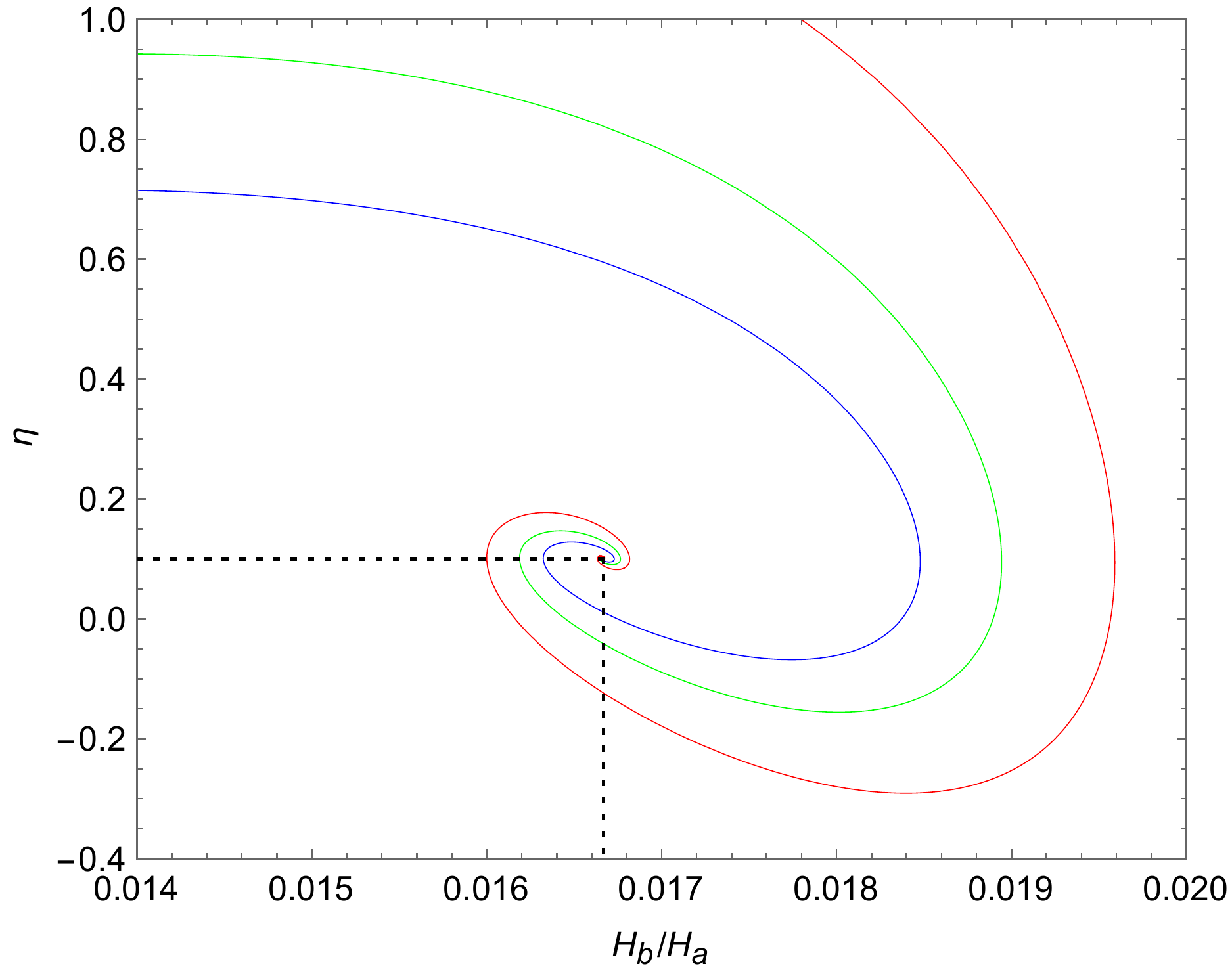}\quad
\includegraphics[scale=0.24]{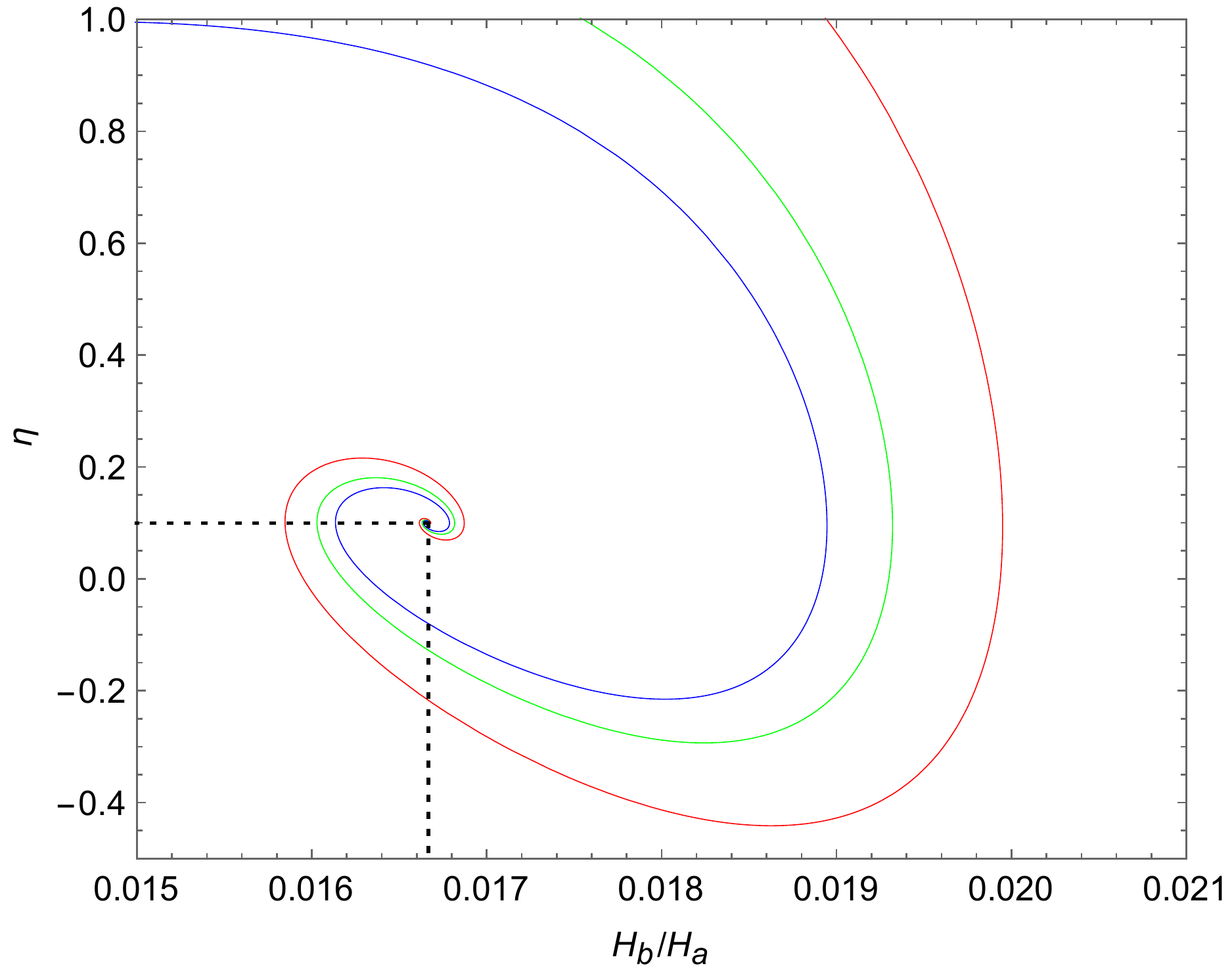}\quad
\includegraphics[scale=0.24]{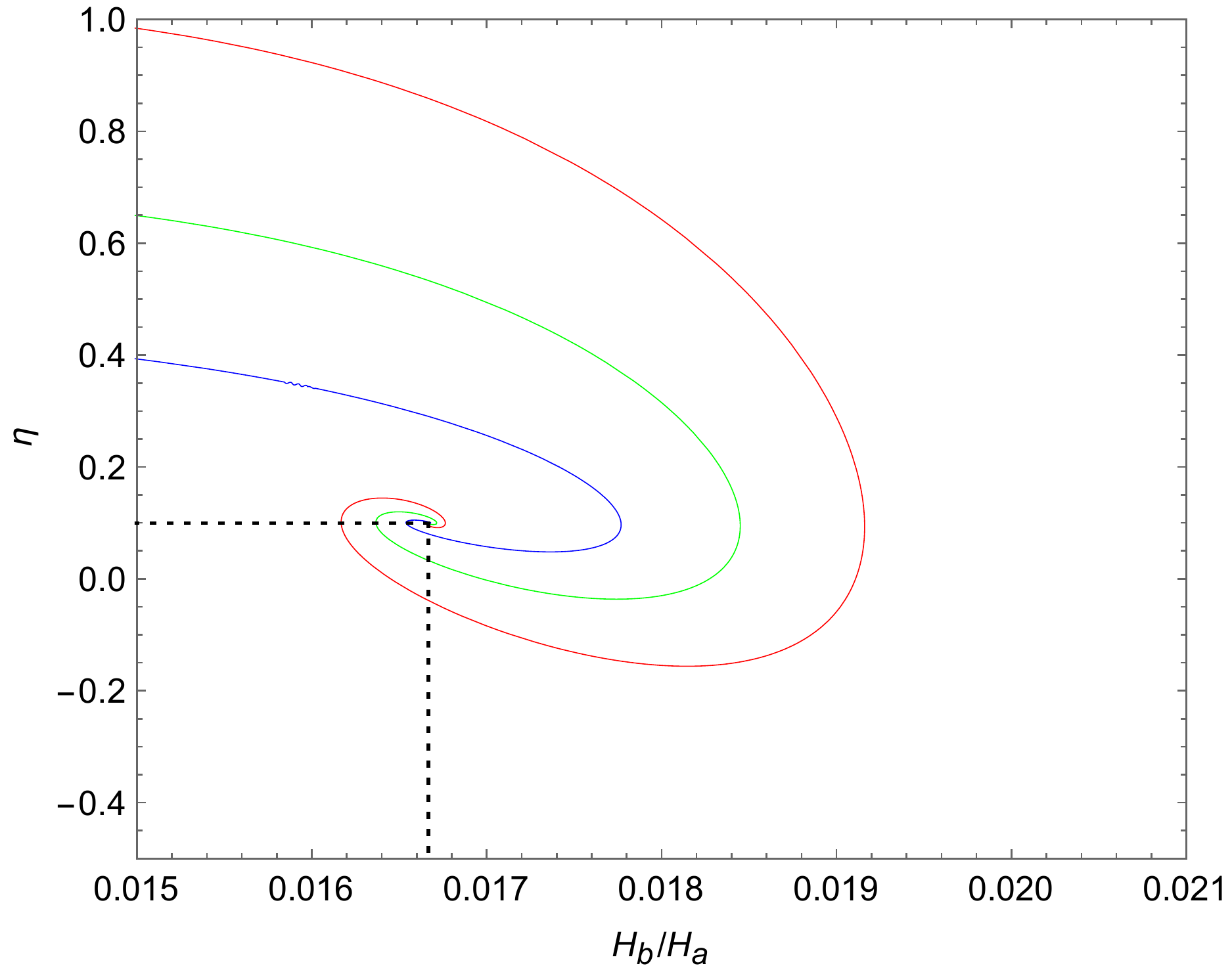}\\
\centering
\caption{Phase space of $H_b(t)/H_a(t)$ and $\eta(t)$ for three different values of $\hat{c}_s$. The left, middle, and right plots correspond to the solutions {I}, {II}, and {III}, respectively. The red, green, and blue curves correspond to $\hat{c}_s=1,~0.95$, and $0.9$, respectively.}
\label{evolution of beta}
\end{figure}

As mentioned above, we will examine in this section the attractive property of  these solutions, following the previous works \cite{Ito:2017bnn,Nguyen:2021emx}. In particular, we will numerically solve Eqs. \eqref{1st Einstein equation}, \eqref{2nd Einstein equation}, \eqref{3rd Einstein equation}, and \eqref{scalar field equation} with the  functions $f(\phi)$, $L(\phi)$, and $V(\phi)$ derived under the constant-roll condition along with the other ones to see the time evolution of the ratio $H_b(t)/H_a(t)$, the speed of sound $c_s(t)$, and the slow-roll parameter $\eta(t)$. It turns out that the obtained solutions will be confirmed to be attractive if $H_b(t)/H_a(t) \to n$, $c_s(t) \to \hat c_s$, and $\eta(t) \to \hat\beta $ when an e-fold number goes to $N\simeq 60$. It should be noted that the field parameters will be chosen as $\beta=0.1$ and $|M|=10^{-5}$ with three different values of $\hat{c}_s=1,~0.95$, and $0.9$. We will also choose initial conditions as $\phi(0)=15$ and $\dot{\phi}(0)=0$. As a result, numerical results presented in Figs. \ref{evolution of n}, \ref{evolution of cs}, and \ref{evolution of beta} clearly indicate that the obtained anisotropic constant-roll inflationary solutions of this {\it k}-inflation model all turn out to be attractive as expected. An interesting point should be noted is that the decrease of $\hat c_s$ tends to delay the convergence of $H_b(t)/H_a(t)$ as well as $c_s(t)$, similar to that obtained in the DBI model \cite{Nguyen:2021emx}. This might be a general feature of non-canonical anisotropic constant-roll inflation models. 

\begin{figure}[hbtp]
\includegraphics[scale=0.35]{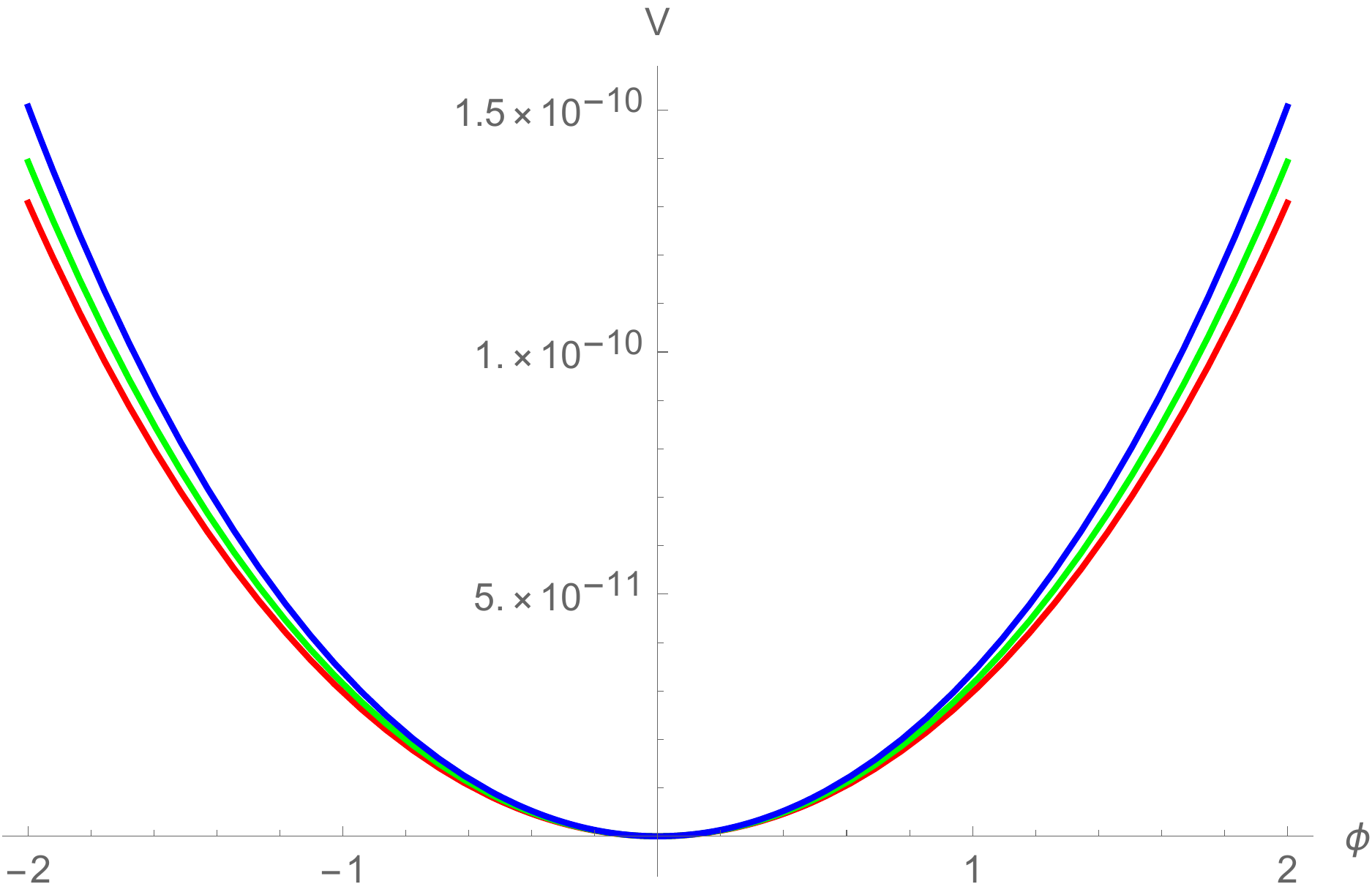}\quad
\includegraphics[scale=0.35]{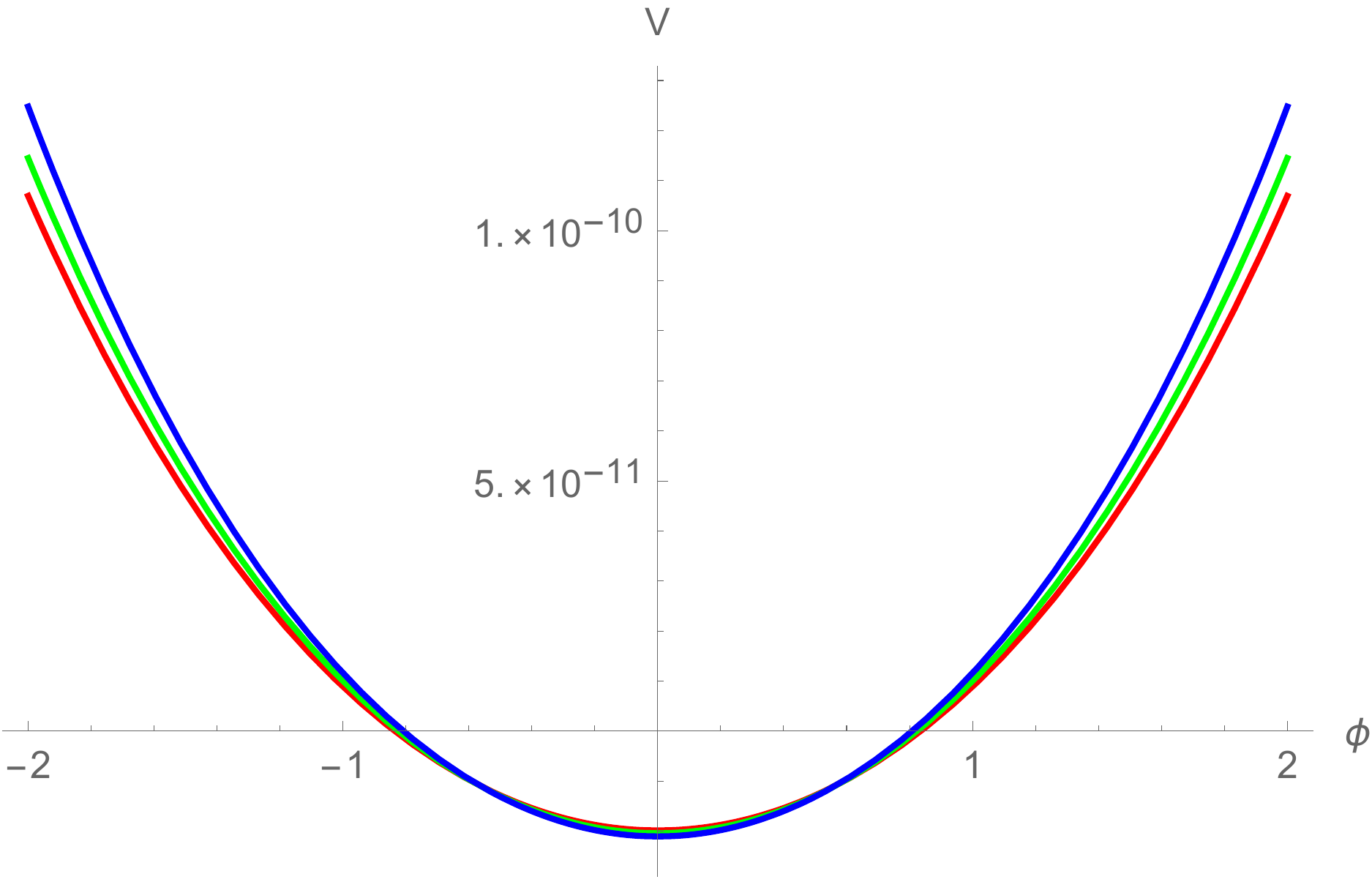}\\
\centering
\caption{Behavior of the potential $V(\phi)$ in Eq. \eqref{V sinh} for the solutions II (left) and III (right) with different values of $\hat{c}_s$. The red, green, and blue curves correspond to $\hat{c}_s=1,~0.95$, and $0.9$, respectively.}
\label{V_lower_sign}
\end{figure}

Before ending this section, we would like to discuss a bit of the difference between the solution III and the solutions I and II, which can be seen in Figs. \ref{evolution of n} and \ref{evolution of cs}. In particular, we observe, according to the number of e-folds, that the inflationary phase of the solution III tends to stop before the e-fold number reaches to $50-60$, in contrast to the solutions I and II. This is due to the fact that the negativity of the potential $V(\phi)$ happens sooner than expected when $\phi$ goes to zero (see Fig. \ref{V_lower_sign} for details). Similar properties can be found in the non-canonical DBI one \cite{Nguyen:2021emx}.

\section{\Mr{Conclusions}} \label{conclusion}
We have investigated whether the {\it k}-inflation model admits anisotropic inflationary solutions under the constant-roll condition in the presence of supergravity-motivated coupling between the scalar and vector fields, i.e., $f^2(\phi)F_{\mu\nu}F^{\mu\nu}$, along with the potential $V(\phi)$.  As a result, the constant-roll condition \eqref{constant-roll condition} together with additional constancy conditions \eqref{constant anisotropy condition} and \eqref{constant speed of sound condition} imply not only the typical power-law type of inflation shown in Eqs. \eqref{a-power} and \eqref{b-power} but also the novel types of inflation shown in Eqs. \eqref{a-upper-sign}, \eqref{b-upper-sign}, \eqref{a-lower-sign}, and \eqref{b-lower-sign}, whose cosmological consequences would be investigated elsewhere. Furthermore, numerical calculations have been performed to confirm that these solutions are indeed attractive during the inflationary phase. This result indicates that the cosmic no-hair conjecture is invalid in this anisotropic constant-roll {\it k}-inflation model. The present work together with the previous ones in Refs. \cite{Ito:2017bnn,Nguyen:2021emx} point out that the cosmic no-hair conjecture seems to be extensively violated in the context of anisotropic constant-roll inflation. Therefore, it is important to investigate whether extensions of this model, e.g., the Gauss-Bonnet extension \cite{Tuan:2019jgv}, favor the cosmic no-hair conjecture. We will leave this issue for our further studies.  

\section*{ACKNOWLEDGMENT}
We would like to thank two anonymous referees very much for their useful comments and suggestions.
D.H.N. was funded by Vingroup Joint Stock Company and supported by the Domestic Master/ PhD Scholarship Programme of Vingroup Innovation Foundation (VINIF), Vingroup Big Data Institute (VINBIGDATA), code  VINIF.2021.ThS.48. T.Q.D. would like to thank Prof. P. V. Dong and Dr. N. T. Duy for their useful helps. T.Q.D. would also like to thank Prof. W. F. Kao for his tireless collaboration on anisotropic inflation. This study is supported by the Vietnam National Foundation for Science and Technology Development (NAFOSTED) under Grant number 103.01-2020.15.  



\begin{thebibliography}{99}  
 \bibitem{WMAP}
  G.~Hinshaw {\it et al.} [WMAP Collaboration],
{\it  Nine-year Wilkinson Microwave Anisotropy Probe (WMAP) observations: Cosmological parameter results}, 
{\it  Astrophys.\ J.\ Suppl.\ }  {\bf 208} (2013) 19 
  [arXiv:1212.5226].

\bibitem{Planck}
Y.~Akrami \textit{et al.} [Planck Collaboration],
{\it Planck 2018 results. VII. Isotropy and statistics of the CMB},
{\it Astron. Astrophys.} \textbf{641} (2020) A7 
[arXiv:1906.02552].
  
\bibitem{cosmological-principle}
  D.~Saadeh, S.~M.~Feeney, A.~Pontzen, H.~V.~Peiris and J.~D.~McEwen,
{\it How isotropic is the Universe?},
 {\it Phys.\ Rev.\ Lett.\ }{\bf 117} (2016) 131302   
  [arXiv:1605.07178];
J. Soltis, A. Farahi, D. Huterer and C. M. Liberato II,
{\it Percent-level test of isotropic expansion using type Ia supernovae},
 {\it Phys. Rev. Lett.\ }{\bf 122} (2019)  091301  
 [arXiv:1902.07189];
N.~J.~Secrest, S.~von Hausegger, M.~Rameez, R.~Mohayaee, S.~Sarkar and J.~Colin,
{\it A test of the cosmological principle with quasars},
{\it Astrophys. J. Lett.} \textbf{908} (2021)  L51 
[arXiv:2009.14826];
C.~Krishnan, R.~Mohayaee, E.~\'O.~Colg\'ain, M.~M.~Sheikh-Jabbari and L.~Yin,
{\it Hints of FLRW breakdown from supernovae},
{\it Phys. Rev. D} \textbf{105} (2022) 063514 
[arXiv:2106.02532].


\bibitem{FLRW}
  T.~Buchert, A.~A.~Coley, H.~Kleinert, B.~F.~Roukema and D.~L.~Wiltshire,
 {\it Observational challenges for the standard FLRW model},
  {\it Int.\ J.\ Mod.\ Phys.\ D} {\bf 25} (2016) 1630007 
  [arXiv:1512.03313].
  
   \bibitem{bianchi}
  G.~F.~R.~Ellis and M.~A.~H.~MacCallum,
 {\it A Class of homogeneous cosmological models},
{\it  Commun.\ Math.\ Phys.\ }  {\bf 12} (1969) 108;
  G.~F.~R.~Ellis,
{\it  The Bianchi models: Then and now},
{\it  Gen.\ Rel.\ Grav.\ } {\bf 38} (2006) 1003.


\bibitem{Schwarz:2015cma}
D.~J.~Schwarz, C.~J.~Copi, D.~Huterer and G.~D.~Starkman,
{\it CMB Anomalies after Planck},
{\it Class. Quant. Grav.} {\bf 33} (2016) 184001 
[arXiv:1510.07929].
  

\bibitem{Krishnan:2021dyb}
C.~Krishnan, R.~Mohayaee, E.~\'O.~Colg\'ain, M.~M.~Sheikh-Jabbari and L.~Yin,
{\it Does Hubble tension signal a breakdown in FLRW cosmology?},
{\it Class. Quant. Grav.} \textbf{38} (2021) 184001 
[arXiv:2105.09790].

 \bibitem{cosmic-inflation} 
A.~A.~Starobinsky,
{\it A new type of isotropic cosmological models without singularity},
{\it Phys. Lett. B} {\bf 91} (1980) 99;
  A.~H.~Guth,
 {\it The inflationary universe: A possible solution to the horizon and flatness problems},
  {\it Phys.\ Rev.\ D} {\bf 23} (1981) 347;
  A.~D.~Linde,
{\it  A new inflationary universe scenario: A possible solution of the horizon, flatness, homogeneity, isotropy and primordial monopole problems},
 {\it Phys.\ Lett.\ B} {\bf 108} (1982) 389;
  A.~D.~Linde,
{\it  Chaotic inflation},
 {\it Phys.\ Lett.\ B} {\bf 129} (1983) 177.

\bibitem{Pitrou:2008gk} 
  C.~Pitrou, T.~S.~Pereira and J.~P.~Uzan,
{\it  Predictions from an anisotropic inflationary era},
 {\it J. Cosmol. Astropart. Phys.} {\bf 04} (2008) 004 
  [arXiv:0801.3596];
  A.~E.~Gumrukcuoglu, C.~R.~Contaldi and M.~Peloso,
{\it  Inflationary perturbations in anisotropic backgrounds and their imprint on the CMB},
{\it J. Cosmol. Astropart. Phys.} {\bf 07} (2007) 005 
  [arXiv:0707.4179].
  
\bibitem{Colin:2018ghy}
J.~Colin, R.~Mohayaee, M.~Rameez and S.~Sarkar,
{\it Evidence for anisotropy of cosmic acceleration},
{\it Astron. Astrophys.} {\bf 631} (2019) L13 
[arXiv:1808.04597].

\bibitem{cosmic-no-hair} 
  G.~W.~Gibbons and S.~W.~Hawking,
  {\it Cosmological event horizons, thermodynamics, and particle creation},
   {\it Phys.\ Rev.\  D} {\bf 15} (1977) 2738;
  S.~W.~Hawking and I.~G.~Moss,
 {\it Supercooled phase transitions in the very early universe},
  {\it Phys.\ Lett.\ B} {\bf 110} (1982) 35.
  
\bibitem{wald} 
  R.~M.~Wald,
  {\it Asymptotic behavior of homogeneous cosmological models in the presence of a positive cosmological constant},
 {\it Phys.\ Rev.\ D} {\bf 28} (1983) 2118.

\bibitem{proof-1}
  J.~D.~Barrow,
  {\it Cosmic no hair theorems and inflation},
 {\it Phys.\ Lett.\ B} {\bf 187} (1987) 12;
M.~Mijic and J.~A.~Stein-Schabes,
{\it A no-hair theorem for $R^{2}$ models},
{\it Phys. Lett. B} \textbf{203} (1988) 353;
  Y.~Kitada and K.~i.~Maeda,
 {\it  Cosmic no hair theorem in power law inflation},
{\it  Phys.\ Rev.\ D} {\bf 45} (1992) 1416.


\bibitem{proof-2}
  M.~Kleban and L.~Senatore,
 {\it Inhomogeneous anisotropic cosmology},
  {\it J. Cosmol. Astropart. Phys.} {\bf 10} (2016) 022
 [arXiv:1602.03520]; 
  W.~E.~East, M.~Kleban, A.~Linde and L.~Senatore,
{\it  Beginning inflation in an inhomogeneous universe},
  {\it J. Cosmol. Astropart. Phys.}  {\bf 09} (2016) 010
[arXiv:1511.05143].

\bibitem{proof-3}
  S.~M.~Carroll and A.~Chatwin-Davies,
{\it  Cosmic equilibration: A holographic no-hair theorem from the generalized second law},
 {\it Phys.\ Rev.\ D} {\bf 97} (2018) 046012 
  [arXiv:1703.09241].
  
\bibitem{Azhar:2022yip}
F.~Azhar and D.~I.~Kaiser,
{\it Flows into de Sitter from anisotropic initial conditions: An effective field theory approach},
{ arXiv:2207.08355}.
 

 \bibitem{local-1}
A.~A.~Starobinsky, 
{\it Isotropization of arbitrary cosmological expansion given an effective cosmological constant}, 
{\it JETP Lett.} \textbf{37} (1983) 66; 
 V.~Muller, H.~J.~Schmidt and A.~A.~Starobinsky, 
 {\it Power law inflation as an attractor solution for inhomogeneous cosmological models},
{\it Class. Quant. Grav.} \textbf{7} (1990) 1163.

\bibitem{local-2}
J.~D.~Barrow and J.~Stein-Schabes,
{\it Inhomogeneous cosmologies with cosmological constant},
{\it Phys. Lett. A} \textbf{103} (1984) 315;
 L.~G.~Jensen and J.~A.~Stein-Schabes,
{\it Is inflation natural?},
{\it Phys. Rev. D} \textbf{35} (1987) 1146;
J.~A.~Stein-Schabes,
{\it Inflation in spherically symmetric inhomogeneous models},
{\it Phys. Rev. D} \textbf{35} (1987) 2345.
 
\bibitem{MW0} 
  M.~a.~Watanabe, S.~Kanno and J.~Soda,
 {\it Inflationary universe with anisotropic hair},
  {\it Phys.\ Rev.\ Lett.\ }  {\bf 102} (2009) 191302 
  [arXiv:0902.2833].

\bibitem{MW} 
  S.~Kanno, J.~Soda and M.~a.~Watanabe,
 {\it  Anisotropic power-law inflation},
 {\it J. Cosmol. Astropart. Phys.} {\bf 12} (2010) 024 
  [arXiv:1010.5307].

\bibitem{extensions}
  R.~Emami, H.~Firouzjahi, S.~M.~Sadegh Movahed and M.~Zarei,
 {\it Anisotropic inflation from charged scalar fields},
  {\it J. Cosmol. Astropart. Phys.} {\bf 02} (2011)  005
  [arXiv:1010.5495];
  K.~Murata and J.~Soda,
  {\it Anisotropic inflation with non-Abelian gauge kinetic function},
  {\it J. Cosmol. Astropart. Phys.} {\bf 06} (2011) 037
  [arXiv:1103.6164];
  S.~Hervik, D.~F.~Mota and M.~Thorsrud,
{\it  Inflation with stable anisotropic hair: is it cosmologically viable?},
  {\it J. High Energy Phys.}  {\bf 11} (2011) 146
  [arXiv:1109.3456];
  M.~Thorsrud, D.~F.~Mota and S.~Hervik,
{\it  Cosmology of a scalar field coupled to matter and an isotropy-violating Maxwell field},
  {\it J. High Energy Phys.} {\bf 10}  (2012) 066
  [arXiv:1205.6261];
  J.~Holland, S.~Kanno and I.~Zavala,
{\it  Anisotropic inflation with derivative couplings},
  {\it Phys.\ Rev.\ D} {\bf 97} (2018)  103534 
  [arXiv:1711.07450];
  T.~Q.~Do and W.~F.~Kao,
  {\it Anisotropic power-law inflation for a conformal-violating Maxwell model},
{\it  Eur.\ Phys.\ J.\ C} {\bf 78} (2018) 360 
  [arXiv:1712.03755];
T.~Q.~Do and W.~F.~Kao,
{\it Anisotropic power-law inflation for a model of two scalar and two vector fields},
{\it Eur. Phys. J. C} \textbf{81} (2021) 525 
[arXiv:2104.14100];
P.~Gao, K.~Takahashi, A.~Ito and J.~Soda,
{\it Cosmic no-hair conjecture and inflation with an SU(3) gauge field},
{\it Phys. Rev. D} \textbf{104} (2021) 103526 
[arXiv:2107.00264];
C.~B.~Chen and J.~Soda,
{\it Anisotropic hyperbolic inflation},
 {\it J. Cosmol. Astropart. Phys.} \textbf{09} (2021) 026 
[arXiv:2106.04813];
T.~Q.~Do and W.~F.~Kao,
{\it Anisotropic hyperbolic inflation for a model of two scalar and two vector fields},
{\it Eur. Phys. J. C} \textbf{82} (2022) 123 
[arXiv:2110.13516];
C.~B.~Chen and J.~Soda,
{\it Geometric structure of multi-form-field isotropic inflation and primordial fluctuations},
{\it J. Cosmol. Astropart. Phys.} \textbf{05} (2022)  029 
[arXiv:2201.03160].

\bibitem{Do:2011zz} 
  T.~Q.~Do and W.~F.~Kao,
 {\it Anisotropic power-law inflation for the Dirac-Born-Infeld theory},
 {\it Phys.\ Rev.\ D} {\bf 84} (2011) 123009.

\bibitem{Ohashi:2013pca} 
  J.~Ohashi, J.~Soda and S.~Tsujikawa,
{\it  Anisotropic power-law k-inflation},
 {\it Phys.\ Rev.\ D} {\bf 88} (2013) 103517
  [arXiv:1310.3053].
  
\bibitem{Do:2020hjf}
T.~Q.~Do,
{\it Stable small spatial hairs in a power-law k-inflation model},
{\it Eur. Phys. J. C} \textbf{81} (2021)  77 
[arXiv:2007.04867].


\bibitem{Ito:2017bnn}
A.~Ito and J.~Soda,
{\it Anisotropic constant-roll inflation},
{\it Eur. Phys. J. C} \textbf{78} (2018) 55   
[arXiv:1710.09701].

\bibitem{Nguyen:2021emx}
D.~H.~Nguyen, T.~M.~Pham and T.~Q.~Do,
{\it Anisotropic constant-roll inflation for the Dirac\textendash{}Born\textendash{}Infeld model},
{\it Eur. Phys. J. C} \textbf{81} (2021) 839 
[arXiv:2107.14115].

\bibitem{Maleknejad:2012fw} 
  A.~Maleknejad, M.~M.~Sheikh-Jabbari and J.~Soda,
 {\it Gauge fields and inflation},
 {\it Phys.\ Rept.\ }  {\bf 528} (2013) 161 
  [arXiv:1212.2921];
  J.~Soda,
 {\it Statistical anisotropy from anisotropic inflation},
  {\it Class.\ Quant.\ Grav.}  {\bf 29} (2012) 083001 
  [arXiv:1201.6434].
  
 \bibitem{Motohashi:2014ppa}
H.~Motohashi, A.~A.~Starobinsky and J.~Yokoyama,
{\it Inflation with a constant rate of roll},
{\it J. Cosmol. Astropart. Phys.} \textbf{09} (2015) 018 
[arXiv:1411.5021].
  
  
\bibitem{constant-roll-extension}  
H.~Motohashi and A.~A.~Starobinsky,
{\it Constant-roll inflation: confrontation with recent observational data},
{\it Europhys. Lett.} \textbf{117} (2017) 39001 
[arXiv:1702.05847];
J.~T.~Galvez Ghersi, A.~Zucca and A.~V.~Frolov,
{\it Observational constraints on constant roll inflation},
{\it J. Cosmol. Astropart. Phys.} \textbf{05} (2019) 030 
[arXiv:1808.01325];
S.~D.~Odintsov and V.~K.~Oikonomou,
{\it Inflationary dynamics with a smooth slow-roll to constant-roll era transition},
{\it J. Cosmol. Astropart. Phys.} \textbf{04} (2017) 041 
[arXiv:1703.02853];
S.~Nojiri, S.~D.~Odintsov and V.~K.~Oikonomou,
{\it Constant-roll inflation in $F(R)$ gravity},
{\it Class. Quant. Grav.} \textbf{34} (2017)  245012 
[arXiv:1704.05945];
H.~Motohashi and A.~A.~Starobinsky,
{\it $f(R)$ constant-roll inflation},
{\it Eur. Phys. J. C} \textbf{77} (2017) 538 
[arXiv:1704.08188];
V.~K.~Oikonomou,
{\it Reheating in constant-roll $F(R)$ gravity},
{\it Mod. Phys. Lett. A}  \textbf{32} (2017) 1750172 
[arXiv:1706.00507];
L.~Anguelova, P.~Suranyi and L.~C.~R.~Wijewardhana,
{\it Systematics of constant roll inflation},
{\it J. Cosmol. Astropart. Phys.} \textbf{02} (2018) 004 
[arXiv:1710.06989];
A.~Karam, L.~Marzola, T.~Pappas, A.~Racioppi and K.~Tamvakis,
{\it Constant-roll (quasi-)linear inflation},
{\it J. Cosmol. Astropart. Phys.} \textbf{05} (2018) 011 
[arXiv:1711.09861];
A.~Mohammadi, K.~Saaidi and H.~Sheikhahmadi,
{\it Constant-roll approach to non-canonical inflation},
{\it Phys. Rev. D} \textbf{100} (2019) 083520 
[arXiv:1803.01715];
A.~Mohammadi, T.~Golanbari and K.~Saaidi,
{\it Observational constraints on DBI constant-roll inflation},
{\it Phys. Dark Univ.} \textbf{27}, 100456 (2020)
[arXiv:1808.07246];
W.~C.~Lin, M.~J.~P.~Morse and W.~H.~Kinney,
{\it Dynamical analysis of attractor behavior in constant roll inflation},
{\it J. Cosmol. Astropart. Phys.} \textbf{09} (2019) 063 
[arXiv:1904.06289];
H.~Motohashi and A.~A.~Starobinsky,
{\it Constant-roll inflation in scalar-tensor gravity},
{\it J. Cosmol. Astropart. Phys.}  \textbf{11} (2019) 025 
[arXiv:1909.10883];
H.~Motohashi, S.~Mukohyama and M.~Oliosi,
{\it Constant roll and primordial black holes},
{\it J. Cosmol. Astropart. Phys.}  \textbf{03} (2020) 002 
[arXiv:1910.13235];
I.~Antoniadis, A.~Lykkas and K.~Tamvakis,
{\it Constant-roll in the Palatini-$R^2$ models},
{\it J. Cosmol. Astropart. Phys.} \textbf{04} (2020) 033 
[arXiv:2002.12681];
V.~K.~Oikonomou and F.~P.~Fronimos,
{\it A nearly massless graviton in Einstein-Gauss-Bonnet inflation with linear coupling implies constant-roll for the scalar field},
{\it Europhys. Lett.} \textbf{131} (2020)  30001 
[arXiv:2007.11915];
T.~J.~Gao,
{\it Gauss\textendash{}Bonnet inflation with a constant rate of roll},
{\it Eur. Phys. J. C} \textbf{80} (2020) 1013 
[arXiv:2008.03976];
M.~Guerrero, D.~Rubiera-Garcia and D.~Saez-Chillon Gomez,
{\it Constant roll inflation in multifield models},
{\it Phys. Rev. D} \textbf{102} (2020) 123528 
[arXiv:2008.07260];
J.~Sadeghi and S.~Noori Gashti,
{\it Anisotropic constant-roll inflation with noncommutative model and swampland conjectures},
{\it Eur. Phys. J. C} \textbf{81} (2021) 301 
[arXiv:2104.00117];
M.~Shokri, J.~Sadeghi, M.~R.~Setare and S.~Capozziello,
{\it Nonminimal coupling inflation with constant slow roll},
{\it Int. J. Mod. Phys. D} \textbf{30} (2021) 2150070 
[arXiv:2104.00596].
M.~Shokri, M.~R.~Setare, S.~Capozziello and J.~Sadeghi,
{\it Constant-roll $f(R)$ inflation compared with Cosmic Microwave Background anisotropies and swampland criteria},
{\it Eur. Phys. J. Plus} \textbf{137} (2022)  639 
[arXiv:2108.00175].


\bibitem{Odintsov:2019ahz}
S.~D.~Odintsov and V.~K.~Oikonomou,
{\it Constant-roll $k$-inflation dynamics},
{\it Class. Quant. Grav.} \textbf{37} (2020) 025003 
[arXiv:1912.00475].


\bibitem{Martin:2012pe}
J.~Martin, H.~Motohashi and T.~Suyama,
{\it Ultra slow-roll inflation and the non-Gaussianity consistency relation},
{\it Phys. Rev. D} \textbf{87} (2013) 023514 
[arXiv:1211.0083].


\bibitem{Abbott:1984fp}
L.~F.~Abbott and M.~B.~Wise,
{\it Constraints on generalized inflationary cosmologies},
{\it Nucl. Phys. B} \textbf{244} (1984) 541;
F.~Lucchin and S.~Matarrese,
{\it Power law inflation},
{\it Phys. Rev. D} \textbf{32} (1985) 1316.

\bibitem{Barrow:1994nt}
J.~D.~Barrow,
{\it Exact inflationary universes with potential minima},
{\it Phys. Rev. D} \textbf{49} (1994) 3055.

\bibitem{Boubekeur:2005zm}
L.~Boubekeur and D.~H.~Lyth,
{\it Hilltop inflation},
{\it J. Cosmol. Astropart. Phys.} \textbf{07} (2005) 010 
[hep-ph/0502047].

  \bibitem{ArmendarizPicon:1999rj} 
  C.~Armendariz-Picon, T.~Damour and V.~F.~Mukhanov,
 {\it $k$-inflation},
 {\it  Phys.\ Lett.\ B} {\bf 458} (1999) 209 
  [hep-th/9904075];
  J.~Garriga and V.~F.~Mukhanov,
 {\it Perturbations in $k$-inflation},
 {\it  Phys.\ Lett.\ B} {\bf 458} (1999) 219 
  [hep-th/9904176].
  
\bibitem{Ellis:2013xoa}
 J.~Ellis, D.~V.~Nanopoulos and K.~A.~Olive,
{\it No-scale supergravity realization of the Starobinsky model of inflation},
{\it Phys. Rev. Lett.} \textbf{111} (2013) 111301 
[erratum: {\it Phys. Rev. Lett.} \textbf{111} (2013) 129902]
[arXiv:1305.1247].

\bibitem{Farakos:2013cqa}
F.~Farakos, A.~Kehagias and A.~Riotto,
{\it On the Starobinsky model of inflation from supergravity},
{\it Nucl. Phys. B} \textbf{876} (2013) 187 
[arXiv:1307.1137].

  
\bibitem{Yamaguchi:2011kg}
 M.~Yamaguchi,
{\it Supergravity based inflation models: a review},
{\it Class. Quant. Grav.} \textbf{28} (2011) 103001 
[arXiv:1101.2488].
  
\bibitem{Tuan:2019jgv}
T. Q. Do and S.~H.~Q.~Nguyen,
{\it No small hairs in anisotropic power-law Gauss-Bonnet inflation},
{\it Commun. in Phys.} \textbf{29} (2019) 173 
[arXiv:1905.01427]; 
T.~M.~Pham, D.~H.~Nguyen and T.~Q.~Do,
{\it k-Gauss-Bonnet inflation},
arXiv:2107.05926.

\end{thebibliography}

\end{document}